\documentclass[sigconf]{acmart}

\usepackage{xcolor}
\definecolor{thedarkblue}{RGB}{0,0,120} %104} % 180
\definecolor{mydarkblue}{rgb}{0,0.08,0.45} %ICML dark blue
\definecolor{darkblue}{rgb}{0,0.08,180}
\colorlet{TufteRed}{red!80!black}
\definecolor{theblue}{RGB}{0,0,180}
\colorlet{thered}{TufteRed}
      
\usepackage{microtype}
\usepackage{balance}

\usepackage{booktabs}
\usepackage{tabularx}

\usepackage{amsmath,amsthm}

\newcommand{\eat}[1]{\ignorespaces}
\usepackage{comment}

\usepackage{tikz}
\usepackage{verbatim}
\usetikzlibrary{arrows}
\usetikzlibrary{shapes,snakes}
\usetikzlibrary{decorations.pathmorphing} % noisy shapes
\usetikzlibrary{fit}					% fitting shapes to coordinates
\usetikzlibrary{backgrounds}	% drawing the background after the foreground

\usepackage{ragged2e}
\usepackage{multirow}
\usepackage{microtype}
\usepackage{balance}
\usepackage{setspace}

\graphicspath{{./}{./graphics/}}
\newcolumntype{H}{>{\setbox0=\hbox\bgroup}c<{\egroup}@{}}

\newcolumntype{R}[1]{>{\RaggedLeft\arraybackslash}} %p{#1}}
\newcolumntype{L}[1]{>{\RaggedRight\arraybackslash}} %p{#1}}

\AtBeginEnvironment{pmatrix}{\setlength{\arraycolsep}{2pt}}

\DeclareMathOperator{\hugeE}{\mbox{\huge\raise-0.3ex\hbox{E}}}
\DeclareMathOperator{\p}{\mathbb{P}}
\DeclareMathOperator{\hugep}{\mbox{\huge\raise-0.3ex\hbox{$\p$}}}

\AtBeginDocument{
  \providecommand\BibTeX{{
    \normalfont B\kern-0.5em{\scshape i\kern-0.25em b}\kern-0.8em\TeX}}}

\copyrightyear{2023} 
\acmYear{2023} 
\setcopyright{rightsretained} 
\acmConference[UIST '23]{The 36th Annual ACM Symposium on User Interface Software and Technology}{October 29-November 1, 2023}{San Francisco, CA, USA}
\acmBooktitle{The 36th Annual ACM Symposium on User Interface Software and Technology (UIST '23), October 29-November 1, 2023, San Francisco, CA, USA}
\acmDOI{10.1145/3586183.3606807}
\acmISBN{979-8-4007-0132-0/23/10}

\begin{document}

\title{TaleStream: Supporting Story Ideation with Trope Knowledge}

\author{Jean-Peïc Chou}
\affiliation{%
  \institution{Stanford University}
  % \city{Stanford}
  % \state{California}
  \country{}
}
% \email{jeanpeic@stanford.edu}

\author{Alexa F. Siu}
\affiliation{%
  \institution{Adobe Research}
  % \city{San Jose}
  % \state{California}
  \country{}
}
% \email{asiu@adobe.com}

\author{Nedim Lipka}
\affiliation{%
  \institution{Adobe Research}
  % \city{San Jose}
  % \state{California}
  \country{}
}
% \email{lipka@adobe.com}

\author{Ryan Rossi}
\affiliation{%
  \institution{Adobe Research}
  % \city{San Jose}
  % \state{California}
  \country{}
}
% \email{ryrossi@adobe.com}

\author{Franck Dernoncourt}
\affiliation{%
  \institution{Adobe Research}
  % \city{Seattle}
  % \state{Washington}
  \country{}
}
% \email{dernonco@adobe.com}

\author{Maneesh Agrawala}
\affiliation{%
  \institution{Stanford University}
  % \city{Stanford}
  % \state{California}
  \country{}
}
\affiliation{%
  \institution{Roblox}
  % \city{San Mateo}
  % \state{California}
  \country{}
}
% \email{maneesh@cs.stanford.edu}

\renewcommand{\shortauthors}{Chou et al.}

\begin{teaserfigure}
\centering
\includegraphics[width=\textwidth]{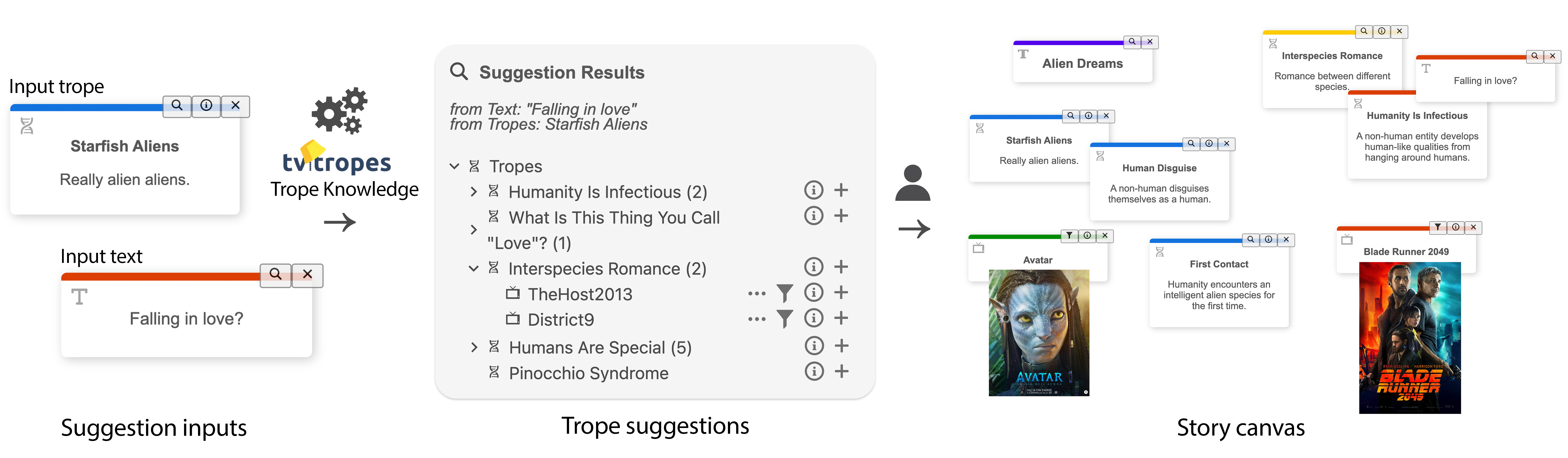}
\caption{Our system allows users to generate story ideas by providing a set of inputs, which can include tropes and text, e.g., the \textit{Starfish Aliens} trope and "Falling in love?" (left). Leveraging trope knowledge extracted from tvtropes.org, our suggestion algorithm automatically surfaces relevant tropes as story ideas with additional information (middle). Users can add some of these suggestions to their canvas, e.g., \textit{Humanity is Infectious} and \textit{Interspecies Romance} (right). Through iteration with the system, users can ideate and develop their stories.}
\label{fig:teaser}
\Description{Overview of TaleStream creation workflow. The figure consists of a sequence of 3 parts with arrows between them from left to right. The Left part shows two input index notes. The middle part shows the results provided by Talestream as a list of tropes. The right part shows many index notes representing the author's story.}
\end{teaserfigure}

\begin{abstract}
Story ideation is a critical part of the story-writing process. It is challenging to support computationally due to its exploratory and subjective nature. Tropes, which are recurring narrative elements across stories, are essential in stories as they shape the structure of narratives and our understanding of them. In this paper, we propose to use tropes as an intermediate representation of stories to approach story ideation. We present TaleStream, a canvas system that uses tropes as building blocks of stories while providing steerable suggestions of story ideas in the form of tropes. Our trope suggestion methods leverage data from the tvtropes.org wiki. We find that 97\% of the time, trope suggestions generated by our methods provide better story ideation materials than random tropes. Our system evaluation suggests that TaleStream can support writers’ creative flow and greatly facilitates story development. Tropes, as a rich lexicon of narratives with available examples, play a key role in TaleStream and hold promise for story-creation support systems.
\end{abstract}





\maketitle

%\begin{sloppypar}

\section{Introduction}

Finding original and engaging ideas to create compelling stories is a challenging task for authors. Efforts in developing writing support systems have focused on continuing stories by producing text blocks to add and edit \cite{roemmele_creative_2015, mirowski_co_2023}. However, beyond sentence generation, previous research has suggested that the main strength and use cases of such tools lie in their suggestive power to help overcome writer's block \cite{calderwood_how_2020, yuan_wordcraft_2022, clark_creative_2018, akoury_storium_2020}. In this regard, the focus of our work is to build a system that supports story writing by providing inspiring materials.

To ideate stories, drawing inspiration from existing stories is essential as authors consciously or unconsciously borrow story elements from each other. Such shared elements have been theorized by structuralists since the early 20th century \cite{propp_morphology_1928}. To develop their own stories, authors often rely on well-known narrative structures such as the \textit{Hero’s Journey}, on common narrative devices creating or resolving conflicts like \textit{Love Triangles}, or on archetypal characters such as the \textit{Diabolical Mastermind}. As defined by the community from the wiki tvtropes.org, such recurring narrative elements belonging to common knowledge correspond to tropes, i.e. narrative concepts that “the audience will recognize and understand instantly”. Since 2004, through debates on the wiki forum and iterative modifications, thousands of enthusiasts have been establishing an extensive list of more than 24,000 tropes to break down all stories. Following the structuralist view, such patterns are helpful guidelines for structuring stories or sources of inspiration to develop them. Besides, as recognizable and predictable components, tropes shape the audience’s experience of stories and are, therefore, key elements for authors to grasp to build effective stories.

In this paper, we propose to leverage tropes as ideation fuel and story framework to support the design of stories. We introduce TaleStream, a story-creation support system that uses tropes as story-building blocks (Figure~\ref{fig:teaser}). To provide inspiring materials, instead of generating sentences, TaleStream suggests ideas in the form of tropes and gives access to related knowledge extracted from the wiki tvtropes.com. As needs, desires, and ideas continuously evolve and vary between authors, TaleStream provides steering controls over the suggestions that can be adapted through the creative process. In this regard, we built suggestion algorithms that were independently evaluated from the system. We found that 97\% of the time, trope suggestions generated by our methods provide better story ideation materials than random tropes. In addition, we evaluated TaleStream with experienced story writers, who found the system to be a helpful and flexible creative assistant. Participants shared their enthusiasm for its unique perspective and ability to navigate the space of narratives. The use of tropes was particularly effective in this context, empowering participants to build the backbone of their stories without fear of lacking inspiration while being aware of existing patterns and representations. This work opens up several leads on using tropes as a framework for interacting with intelligent story-creation systems. In summary, this paper makes the following contributions:
\begin{itemize}
    \item TaleStream, a story creation system that emphasizes story ideation by using tropes as building blocks
    \item Two approaches to suggest tropes by leveraging online data and methods to steer the suggestion results with various controls
    \item Results from a summative user study outlining the benefits and limitations of tropes for building stories and story creation tools.
\end{itemize}

\section{Related Work} \label{sec:related-work}

\subsection{Story frameworks}

% Story grammars and frameworks
% Lead by Narrative Structures
Story frameworks were first theorized in ancient Greece by Aristotle \cite{aristotle_poetics_2006}, who argued that tragedies should follow a three-act structure including a beginning, a middle, and an end. In the mid-20th century, structuralist literary theorists regained interest in narrative structures, arguing that all stories shared universal elements and could be essentially reduced to a few narratives. Vladimir Propp’s \textit{Morphology of the Folktale} was a pioneering effort to identify and classify the common narrative elements, or \textit{functions}, of Russian fairy tales \cite{propp_morphology_1928}. Other works have striven to propose more general story frameworks, breaking down stories into a unique storyline, the \textit{Hero's Journey} \cite{campbell_hero_2008}, seven basic plots \cite{booker_seven_2004}, or a more exhaustive list of 1,462 plots \cite{cook_plotto_2011} for instance. The underlying mechanisms aggregating and structuring those elements were theorized in story grammars \cite{lakoff_structural_1972, rumelhart_notes_1975}. Such plot structures are well-known and commonly used by professional story writers. Some of them have been integrated into computational tools to generate stories \cite{gervas_story_2005, grasbon_morphological_2001, fairclough_multiplayer_2003}. For instance, Gervas et al. make use of Propp's elements to develop a story knowledge named \textit{ProppOnto} \cite{gervas_story_2005}. Such approaches correspond to case-based reasoning techniques, which adapt stored stories as frameworks to new contexts. In \cite{swanson_say_2012} and \cite{roemmele_creative_2015}, sentences are directly fetched from a corpus of stories to propose story continuation. Other methods generate stories by adapting higher-level attributes that can consist of goals and events as story units \cite{riedl_vignette-based_2009, turner_minstrel_nodate}. Akoury et al. derive a large corpus of story components from STORIUM, an online collaborative game that lets users write stories based on cards as the framework \cite{akoury_storium_2020}. With TaleStream, we leverage tropes from the wiki tvtrope.org as building blocks of stories. As the result of efforts that have spanned since 2004 by a large community, the more than 24,000 tropes arguably form a comprehensive, organized, and recognizable lexicon for storytelling. Besides, tvtropes.org provide rich information that we propose to make accessible in our system and to reason over for providing suggestions.

\subsection{Tropes}

Since the creation of tvtropes.org, the wiki's data has been extracted several times and made available \cite{kiesel_dbtropes-linked_2010, garcia-ortega_overview_2018} to help people build content generation tools or recommender systems \cite{johnson_scaling_2018}. Many works have proposed global analyses of the website's rich information \cite{mellina_trope_2011, garcia-ortega_tropes_2020, garcia-sanchez_simpsons_2021, chou_structures_2021}. Garcia-Ortega et al. highlighted key statistics about tvtropes.org data \cite{garcia-ortega_tropes_2020} and analyzed the trope co-occurrences in movies to determine a classification of tropes \cite{garcia-sanchez_simpsons_2021}. Chou et al. directly examined the website structure to build a trope-based knowledge graph of storytelling that provide semantic relationships \cite{chou_structures_2021}. Tropes have been shown to be representative of their works, being indicative of their genre \cite{smith_harnessing_2017, aijala_using_2020, datta_latent_2017} or inducing a character's persona \cite{bamman_learning_2013}. Systems such as TropeTwist \cite{alvarez_tropetwist_2022}, Story Designer \cite{alvarez_story_2022}, Ghost \cite{guarneri_ghost_2017}, dairector \cite{eger_dairector_2018}, Dear Leader’s Happy Story Time \cite{horswill_dear_2021}, or StarTroper \cite{garcia-ortega_startroper_2020} proposed to conceive stories with tropes as building blocks, plot points, or narrative beats. These systems were partly inspired by artist James Harris' \textit{Periodic Table of Storytelling}, which draws a comparison between stories, built on tropes, and molecules, composed of atoms \cite{harris_periodic_2017}. For instance, Alavarez et al. design evolutionary narrative structures as graphs with tropes as nodes \cite{alvarez_story_2022, alvarez_tropetwist_2022}. Garcia-Ortega et al. aim at generating lists of tropes by predicting additional tropes that optimize the final story rating \cite{garcia-ortega_startroper_2020}. In dairector, improvisers directly interact with the system that proposes tropes as constraints based on prompts and other plot points. However, these works make use of a limited repertoire of tropes, generally consisting of handpicked ones such as plot tropes or the most recurring ones. This limits users to flesh out stories, whereas we provide a more exhaustive list of tropes ranging from plots to low-level descriptions in TaleStream. In these works, users don't control the design of the story, whereas TaleStream allows users to steer the story idea suggestions. In addition, our work proposes to further the analysis of tropes as structural elements and their use in practice.

\subsection{Story assistants}

Available story assistant tools are numerous. Commercial tools such as \textit{Dramatica} or \textit{Plottr} help authors structure their plot or narrative in their creative workflow \cite{noauthor_dramatica_nodate, noauthor_plottr_nodate}. More recently, there has been a surge of commercial tools leveraging progress with language models \cite{noauthor_sudowrite_nodate, noauthor_ai_nodate}. Besides crowdsourcing efforts to help the creation of stories \cite{kim_mechanical_2017, huang_heteroglossia_2020}, story assistants largely rely on autonomous story generation. Such methods include computational planning, where computers make decisions based on a set of predefined rules and objectives \cite{lebowitz_creating_1984, meehan_tale-spin_1977, riedl_narrative_2010},  character-based simulation \cite{cavazza_characters_2001}, or case-based reasoning \cite{perez_mexica_2001, gervas_story_2005, riedl_vignette-based_2009, riedl_narrative_2010, turner_minstrel_nodate}. Most popular methods now involve language models that infill sentences \cite{ammanabrolu_automated_2020, huang_inset_2020, ippolito_unsupervised_2019, wang_narrative_2020}. Such generative tools, however, necessitate the right level of control \cite{clark_creative_2018}. While some have proposed to guide the generation with keywords \cite{fan_hierarchical_2018, ippolito_unsupervised_2019, sun_iga_2021, xu_megatron-cntrl_2020}, other works have proposed to use natural language prompts directly \cite{duval_breaking_2021}. General conversational agents like ChatGPT \cite{noauthor_introducing_nodate} enable users to refine story generation through an iterative process and are now used by a wider audience \cite{gonsalves_using_2023}. Beyond text, visual elements have also been used as input to control the generation of stories as text \cite{chung_talebrush_2022}. However, the adoption of AI support systems for story creation has raised concerns among professional authors when it comes to translating ideas into words \cite{biermann_from_2022}. With TaleStream, we focus on supporting the ideation part of the story creation process rather than the writing. Instead of fully-fleshed linear sentences, our system relies on tropes for suggestions and as building blocks editable from a canvas. As a structured lexicon built on numerous references, tropes enable authors to navigate the space of existing stories to explore and analyze the proposed ideas.

\section{Talestream: Workflow} \label{sec:workflow}

\begin{figure*} 
\centering
\includegraphics[width=1\linewidth]{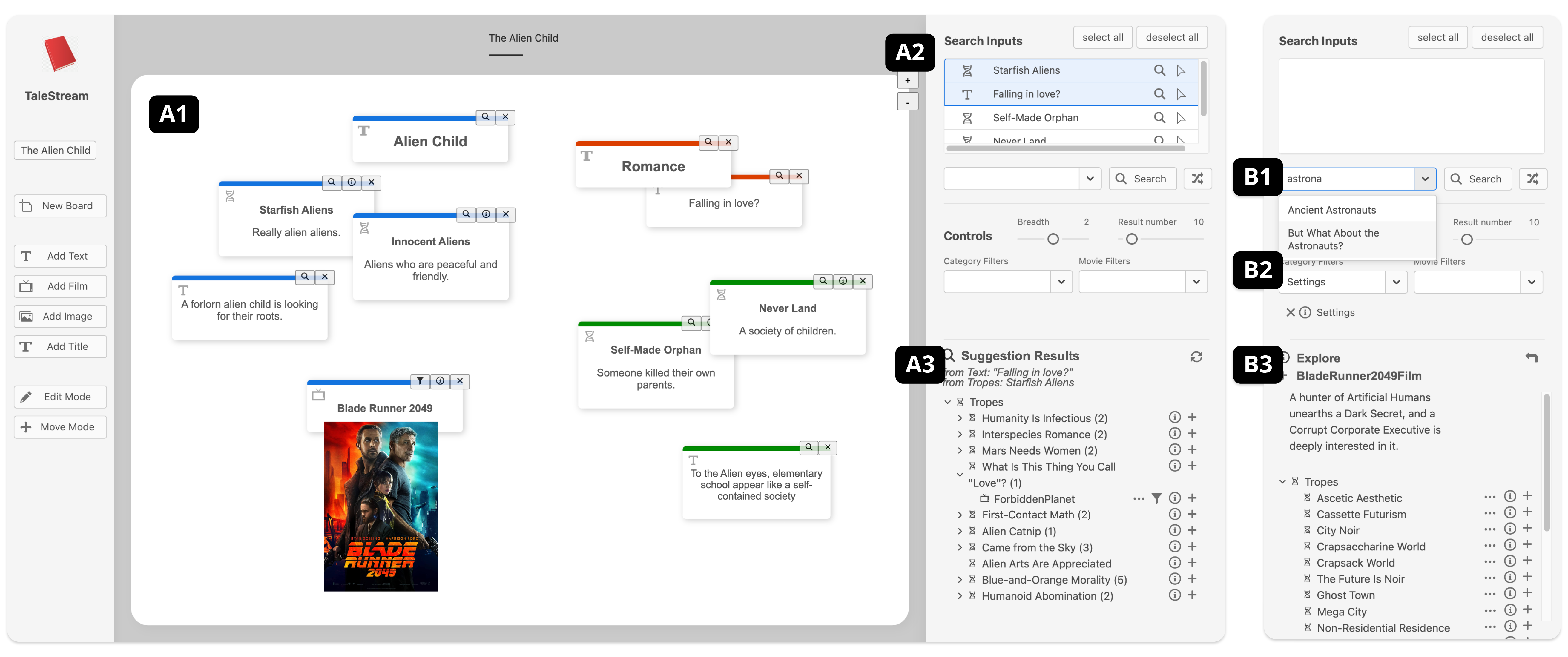}
\caption{TaleStream interface. The Canvas (A1) contains the story elements that can be organized and edited. The Panel List (A2) enables the users to select the canvas elements for consideration when generating trope suggestions. Trope Suggestion Results (A3) include occurrence examples and additional descriptions. Users can use the Text Search Bar (B1) as an additional text entry or to find specific tropes. Additional Controls (B2) allow them to refine the suggestions by specifying their breadth or filtering trope lists. The Explore mode (B3), accessible from information icons, provides access to information on tropes and movies. For example, users can view a list of the \textit{Settings} tropes used in \textit{Blade Runner 2049}.
}
\label{fig:interface}
\Description{The full TaleStream interface is shown on the left. It comprises three main columns. On the left, a grey sidebar with editing tools. In the middle, a white canvas with colorful index notes on it. On the right, a grey sidebar with suggestion controls on top and results at the bottom. This sidebar is shown on the right of the figure but in a different configuration.}
\end{figure*}

We designed TaleStream by deriving from insights and recommendations of past works studying intelligent story writing support tools \cite{calderwood_how_2020, yuan_wordcraft_2022}, as well as from informal interviews that we conducted with five professional story writers. These interviews were aimed at gaining a deeper understanding of their creative processes during story ideation and their experiences with existing intelligent tools. We followed these design guidelines:
\begin{itemize}
    \item DG1: The system should focus on suggesting ideas and encourage users to adapt the generated results to suit their own story
    \item DG2: The system should provide suggestions tailored to the needs of the author throughout the creative process
    \item DG3: The system should provide many suggestions that can be easily replaced and updated.
\end{itemize}
Our TaleStream interface is designed to help authors build their stories by suggesting adapted story ideas in the form of tropes and by giving them access to resources about these narrative patterns. Tropes, as suggestions, provide specific storytelling devices that can be uniquely fitted into the user's story (DG1). The system was iteratively improved with feedback from the five professional story writers through one pilot study. To demonstrate how TaleStream works in practice, we describe the workflow an author might experience to create a story. Our example shows an author developing a story about an “Alien Child”, wishing to incorporate some “Romance” elements, and inspired by the movie \textit{Blade Runner 2049}. As shown in Figure~\ref{fig:interface}, the TaleStream interface consists of three main components: a Canvas (A1), a Control Section (A2, B1, B2), and a Results Section (A3, B3).

\subsection{Board}

The left pane of the interface shows the author’s creation progression as a canvas containing their added story elements (Figure~\ref{fig:interface} A1). The canvas is a drag-and-drop interface in which users add and edit index card elements of different types (Trope, Text, Movie, Title, Image). As a brainstorming tool, 2D canvas interfaces offer flexibility for exploring, visualizing, and organizing story elements and complex ideas. Canvases are widely used in ideation processes across different fields and can be particularly effective when integrated with intelligent agents organizing, retrieving, and suggesting content \cite{koch_may_2019, koch_imagesense_2020, koch_semanticcollage_2020} (DG1). In our example, the canvas was populated by tropes related to a forlorn “Alien Child”, a few text cards describing the story in more detail, and a picture of \textit{Blade Runner 2049} as inspiring reference.

\subsection{Suggestion controls}
During the creation process, users can ask for specific suggestions from TaleStream to develop their stories (DG2). The top-right part of the interface lets authors steer the suggestions based on their creative needs. 

\subsubsection{User Inputs}
Users can specify the inputs of the suggestion results by selecting the canvas tropes and text elements, which can be found in the panel list (Figure~\ref{fig:interface} A2). We found through our pilot study that users wanted to use both trope and text elements to steer the suggestions. In our example, the author can look for ideas to flesh out the \textit{Starfish Aliens} trope and add the text “falling in love?” as input in the search box. An additional text search box is implemented as a combo box containing all the tropes to facilitate the search for tropes – typing ‘astronaut’ in the search bar directly gives the tropes that include ‘astronaut’ in their names (Figure~\ref{fig:interface} B1).

\subsubsection{Filters}
We allow users to specify their search with category and movie filters (Figure~\ref{fig:interface} B2). Categories can notably be used to filter the resulting tropes by narrative function (e.g., Characters, Settings, Beginning Tropes), by theme (e.g., Comedy Tropes, Love Tropes), or by super-trope (e.g., Anti-Hero). With movie filters, users can also get trope suggestions from specific works and get direct inspiration from their content.

\subsubsection{Number of suggestions}
Participants from our pilot study reported that too many suggestions were cognitively overloading. We, therefore, let authors choose the number of displayed suggestions.

\subsection{Results}

\subsubsection{Suggestions}
When asking for suggestions, The Results section of the interface shows a list of trope suggestions (DG3). The author has access to laconic descriptions when hovering over the information icon and can add the corresponding tropes by clicking on the plus button. If a resulting trope co-occurs in movies with at least one of the selected input tropes, up to five of these movies are also displayed, along with a description of how the suggested trope is used in each movie when hovering over the three dots icon. A recap of all the specified inputs is displayed above the results. Responding to the example author’s inputs, the system notably suggests the \textit{Interspecies Romance} (Romance between different species) and \textit{Humanity is Infectious} (A non-human entity develops human-like qualities from hanging around humans) tropes which the author adds to the canvas (Figure~\ref{fig:interface} A3).

\subsubsection{Explore}
Authors can also learn more about a specific movie or trope by clicking on the information icons next to them. This switches the Results section to an Explore mode giving additional information about the element. If the element is a trope, authors have access to the categories it belongs to, its sub-tropes, and the movies it appears in along with the description of how it is implemented in them. If it is a movie, the system displays its synopsis and a complete list of its annotated tropes along with their description. These lists of tropes found in the Explore section can be filtered as well, as shown in the example Figure~\ref{fig:interface} (B3), where the author examines the \textit{Settings} tropes in \textit{Blade Runner 2049} to break down and imagine a similar dark, futuristic, urban, and ascetic atmosphere.

\section{Talestream: Technical details} \label{sec:technical}

In this section, we focus on the technical details of surfacing relevant story ideas and the control mechanisms. Given some inputs from the user, we generate suggestions in the form of tropes that can be used to develop and revise the story. Our methods are based on data extracted from tvtropes.org (Section \ref{sec:tropedata}). These inputs can be of different kinds, either a set of tropes (Section \ref{sec:tropesuggestion}) or some free-form text (Section \ref{sec:tropesearch}), and can be jointly used with additional controls on the generated results (Section \ref{sec:controls}).

\subsection{Trope and movie data} \label{sec:tropedata}

\begin{figure} 
\centering
\includegraphics[width=0.9\linewidth]{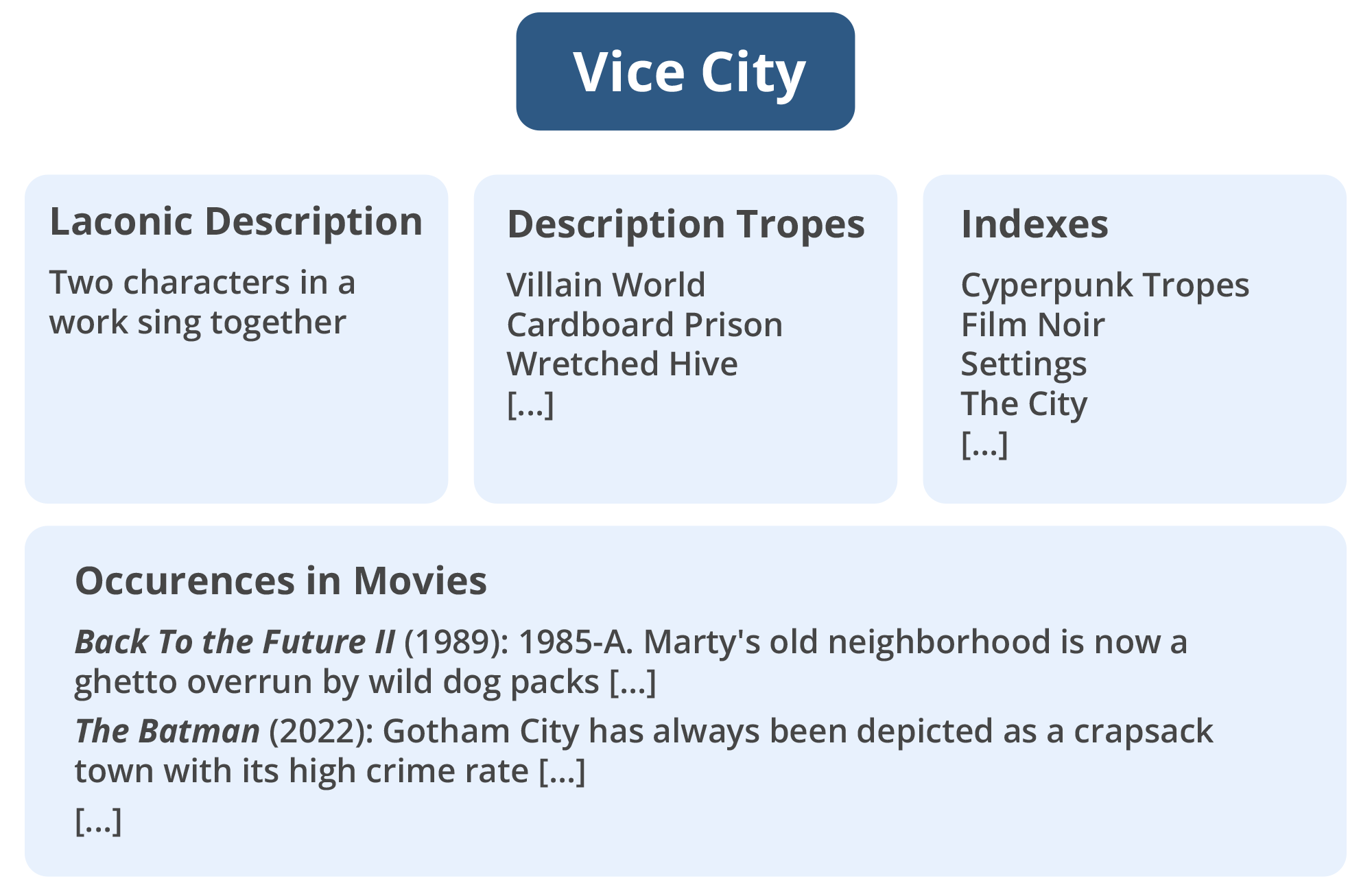}
\caption{Information about the \textit{Vice City} trope extracted from tvtropes.org}
\label{fig:tropeinfo}
\Description{Five blue boxes containing information about the trope title (darker blue), its laconic description, its description tropes, its indexes, and its occurrences In movies.}
\end{figure}

Our methods are based on data extracted from tvtropes.org, a wiki-like website on which a community of enthusiast “tropers” defined more than 24,000 tropes with rich information, as shown in Figure~\ref{fig:tropeinfo}. On the website, tropes are described by a complete description and a “laconic” one. They are organized and grouped by what the tropers name "indexes", i.e. categories, which can be tropes themselves. For instance, the \textit{Anti-Hero} trope is an index that encompasses multiple sub-tropes such as the \textit{Byronic Hero}, the \textit{Justified Criminal}, or the \textit{Moral Sociopath} tropes. Tropes can belong to multiple indexes, which typically group tropes according to a narrative function, theme, genre, medium, or by semantic similarity. The smallest indexes only include 1 trope, e.g., the index \textit{Slice of Life}, which only contains the \textit{Slice of Life Webcomics} trope, while the largest, the index \textit{Comedy Tropes}, includes 1,870 tropes. In addition, the community has annotated trope occurrences in movies and diverse media ranging from literature to advertisement. For each of these occurrences, tropers include a description of the trope implementation in the corresponding work.

Following Chou et al.’s work \cite{chou_structures_2021}, we collect the wiki's tropes and their attributes: their laconic definition, the tropes linked in their descriptions (that we will reference as \textit{description} tropes), their indexes, as well as the movies in which they occur with their implementation details. Table 1 gives an overview of the retrieved dataset. We additionally make use of the MovieLens dataset~\cite{harper_movielens_2015} to provide complementary information about movies. This extracted information is used in our suggestion methods described in the next sections and made available in the Explore section of TaleStream.

\begin{table}[h!]%1
    \centering
     \caption{Overview of the retrieved dataset: Number of extracted elements (left) and tropes' attributes' mean number (right).}
     \label{table:tropekg1}  
     \small
\begin{tabular}{cc}
    \begin{tabular}{ccc}
    \hline
    & Number\\
    \hline
    Tropes & 23,665\\
    Indexes & 1,988\\
    Movies & 15,304\\
    \hline
    % \label{tab:tropekg1}
    \end{tabular}
    &
    \begin{tabular}{ccc}
    \hline
    By trope&Mean number\\
    \hline
    Description tropes & 13.1\\
    Indexes & 4.2\\
    Occurrences & 26.2\\
    \hline
    \end{tabular}
\end{tabular}
    \Description{Two tables are shown. On the left, numbers of Tropes, Indexes, and Movies. On the right, mean number of Description tropes, Indexes, and Occurrences by trope.}
\end{table}

\subsection{Trope suggestion} \label{sec:tropesuggestion}

Our goal is to assist writers by suggesting story ideas in the form of tropes based on a set of input tropes. While prior studies have explored the use of story generation systems to suggest story ideas, these efforts have mainly focused on evaluating text-based features such as grammar, fluency, or lexical cohesion \cite{roemmele_evaluating_2017, purdy_predicting_2018}. The task of generating compelling story ideas is difficult to formulate and evaluate because it is inherently subjective. Our work focuses on ensuring coherence, a key characteristic that has been identified and extensively used in previous literature \cite{alhussain_automatic_2021, sagarkar_quality_2018, castricato_fabula_2021, akoury_storium_2020}. The suggested story ideas should fit seamlessly within the user's narrative, i.e. should be logically consistent with the input tropes.

Our proposed algorithms address the classic trade-off between exploitation and exploration in creativity \cite{boden_creative_2004} and recommender systems \cite{barraza-urbina_exploration-exploitation_2017}. The algorithms are designed to provide suggestions similar to the user's inputs or introduce options that may be less related to broaden their horizons. By balancing these two approaches, our algorithms aim to offer a more comprehensive and personalized suggestion experience for users.

\subsubsection{Index-based method}

Our first method leverages the indexation of the tropes from tvtropes.org to provide coherent and closely related suggestions. With this approach, we focus on suggesting tropes that share similarities (e.g., theme, genre, function) with the input tropes, i.e., propose an exploitation method to retrieve suggestions. Similar tropes can encourage authors to imagine how to refine, combine, and develop the inputs. For instance, based on the input trope \textit{Vice City} (an urban town infested with crimes), the trope \textit{Crapsaccharine World} (a dystopian and grim place disguised in a wonderland) would be an output that shares similarities — both describe a place where darkness and terror reign — and that could be used to develop the input directly.

\begin{figure} 
\centering
\includegraphics[width=1\linewidth]{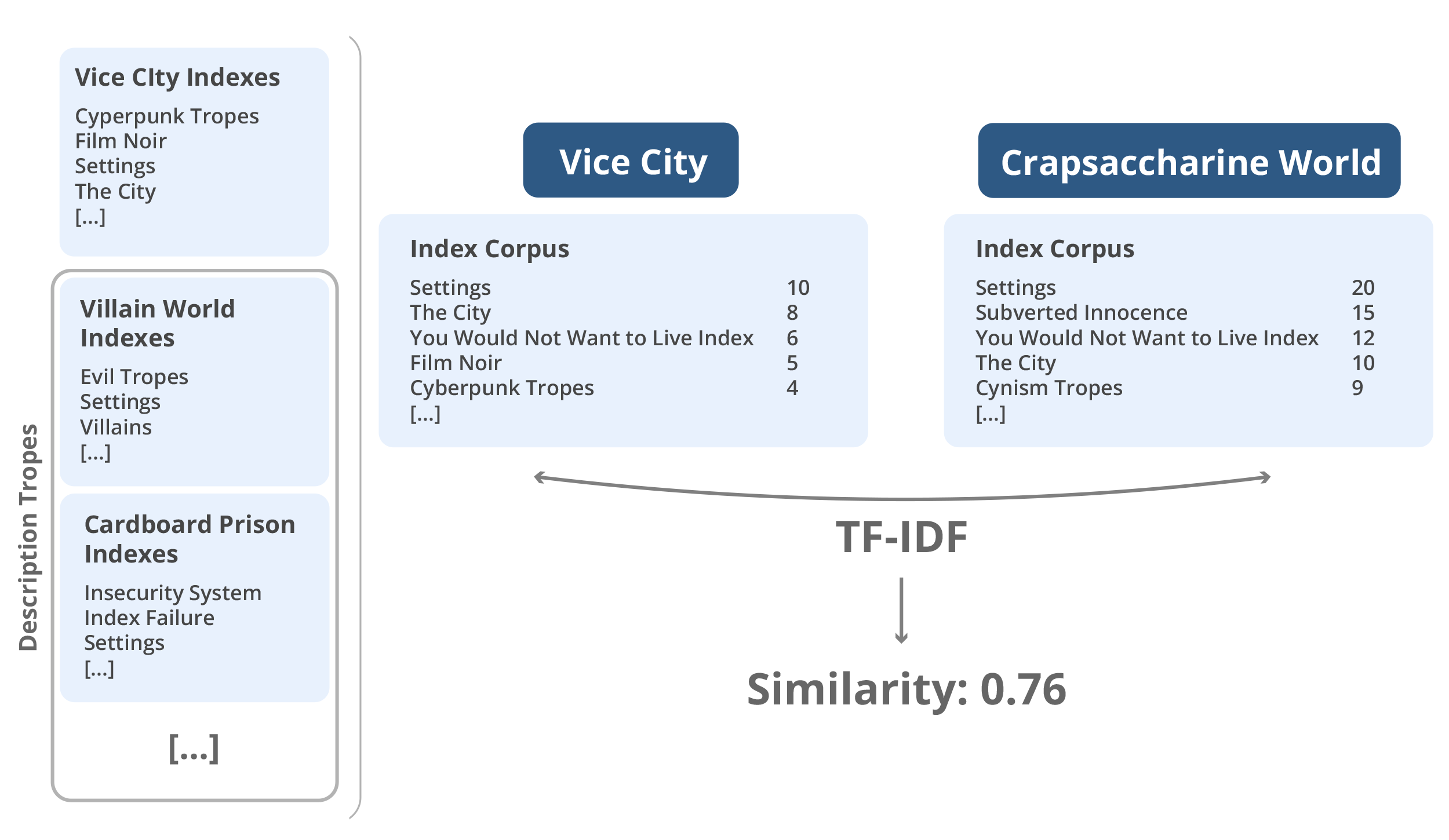}
\caption{Index-based method for single inputs. Indexes of \textit{Vice City} description tropes are aggregated to obtain a broader and weighted Index Corpus. These corpora are then used to compute similarities between tropes with TF-IDF.}
\label{fig:method1}
\Description{On the left, three blue boxes on top of each other contain the trope's indexes and its description tropes' indexes. A bar encompasses them to show that they are aggregated into the considered trope's final Index Corpus. On the right, two blue boxes with two compared tropes' Index Corpus (Index and respective weights in two columns). Underneath them, we can read TF-IDF and an arrow starting from it that points toward "Similarity: 0.76".}
\end{figure}

We compute similarities by comparing tropes based on their annotated indexes. For that, we use sklearn TF-IDF \cite{scikit-learn} by considering tropes as documents and categories as terms. We obtain a corpus of countable indexes for each trope by concatenating the indexes of the trope itself, as well as those of its description tropes, as shown in Figure~\ref{fig:method1}. We use this larger corpus instead of the corpus composed of the trope's indexes only for two main reasons. First, it allows us to weight indexes based on their frequency of occurrence, rather than treating them all equally. This weighting corresponds to having a variable Term Frequency in TF-IDF. Second, this corpus provides more detailed and nuanced information about a trope that the first-order categories may not capture fully (e.g., \textit{Index Failure}, \textit{Cynism Tropes}, \textit{Horror Tropes}). We compute a similarity score between all tropes and the input trope to determine the ones to suggest. Table~\ref{tab:recs} shows the tropes with the highest scores when compared to \textit{Vice City} as the input. For multiple trope inputs, we calculate the final score of each trope by multiplying the similarity scores obtained for each input, thus favoring tropes that are relevant in all aspects:
$$ s_{ind}(\mathcal{E}_{T_i}, T) = \Pi _{T_i \in \mathcal{E}_{T_i}} s_{ind}(T_i, T) $$
where $\mathcal{E}_{T_i}$ is the set of input tropes, $T$ is another trope we compare to, and $s_{ind}$ is the index-similarity scoring function based on sklearn TF-IDF.

\subsubsection{Co-occurrence-based method}

Our second approach relies on trope occurrences in movies. Tropes that often appear in the same movies together are likely to fit easily into the same story. Unlike the first method, this approach doesn't necessarily provide tropes that are close in terms of index, i.e. semantic category. The co-occurrence algorithm captures associations between tropes that may not be obvious or direct, resulting in broader and more "exploratory" suggestions. The method is similar to the previous one. We apply TF-IDF, considering tropes as documents, and movies as terms. We consider the list of movies in which a trope appears as its corpus to compute co-occurrence-similarities between tropes.

With this method, the number of directly co-occurring tropes can be limited, restraining the output coverage. This limitation occurs when a trope appears in only a few movies or in movies that don't have many tropes listed. It poses two problems. Firstly, the suggested tropes are more likely to come from the same story, leading to less diverse and imaginative suggestions. Secondly, tropes that are only compared to a limited number of others are less likely to be suggested overall.

To address this limitation, we also make use of the description tropes, i.e., tropes mentioned in the description of the input trope. For each description trope, we calculate all TF-IDF co-occurrence scores and multiply them by the description trope index-similarity to the input trope to weigh their contribution. Instead of multiplying the scores, which would favor over-represented tropes that appear in all works (e.g., \textit{Big Bad}, \textit{Shout-Out}, \textit{Oh, Crap!}), we keep the maximum score among the computed scores to aggregate the results:
$$ s_{co}(T_i, T) = max_{T_d \in \mathcal{D}_i \cup \{T_i\}}(s_{ind}(T_i, T_d) * \tilde{s}_{co}(T_d, T)) $$
where $T_i$ is the input trope, $T$ is another trope we compare to, $\mathcal{D}_i$ is the set of description tropes of the input, $s_{co}$ and $s_{cat}$ are the category-similarity and co-occurrence-similarity functions, and $\tilde{s}_{co}$ is the first-order co-occurrence-similarity function.

Table~\ref{tab:recs} shows the method results for \textit{Vice City} as input. For multiple input tropes, we calculate the similarity scores for each trope and keep the maximum score among them for the same reason as previously:
$$ s_{co}(\mathcal{E}_{T_i}, T) = max_{T_i \in \mathcal{E}_{T_i}}(s_{co}(T_i, T)) $$

\begin{table}%2
    \centering
    \caption{Tropes with the highest similarity to \textit{Vice City} based on our methods.}
    \label{table:data-stats-node-types}
    \vspace{-2mm}
    \scriptsize
    \begin{tabular}{lll}
    \hline
    \textbf{Index} & \textbf{Co-occurrence} & \textbf{Mixed methods}
    \\
    \hline
      Wretched Hive & False Utopia & City Noir\\ [0,04cm]
    City Noir & Future Society, Present... & Wretched Hive \\ [0,04cm]
    The Big Rotten Apple & Terror Hero & City of Adventure\\ [0,04cm]
    The City & Cataclysm Backstory & Soiled City on a Hill \\ [0,04cm]
    Crapsaccharine World & Color-Coded Castes & City on a Bottle \\ [0,02cm]
     \hline
    \end{tabular}
    \label{tab:recs}
    \Description{Three columns of trope suggestion results. The columns' titles correspond to the methods: Index, Co-occurrence, and Mixed methods.}
\end{table}

\subsection{Trope search} \label{sec:tropesearch}

We allow users to obtain story ideas from plain text. To implement this feature, we look for the tropes that are the most similar to the input text. We once again use sklearn TF-IDF. Tropes are still the documents, and we use the examples extracted from the website movie pages to compose their corpus. Each trope corpus is obtained by concatenating its implementation descriptions in movies.

\subsection{Suggestion controls} \label{sec:controls}

\subsubsection{Breadth}

We provide authors with a Breadth slider feature that controls the method to use to let them select the desired degree of exploitation versus exploration. Authors can set the Breadth slider to 1 to use the index-based method and to 3 for the co-occurrence method.  For a balanced approach, authors can set the slider to 2, which combines both methods by multiplying their scores. A result example of the mixed method is shown in Table~\ref{tab:recs}. To help users understand the connections between the input and output tropes, we provide examples of movies in which both appear together.

\subsubsection{Mixing search inputs}

To combine suggestions based on both trope and text inputs, output scores from the trope inputs are multiplied by the ones from the text inputs. This ensures that the combined result prioritizes items that are relevant to both the trope and text queries.

$$ \tilde{s}(\mathcal{E}_{T_i}, T) = s_{trope}(\mathcal{E}_{T_i}, T) * s_{text}(text) $$
where $s_{trope}$ is the tropes-to-trope similarity function, and $s_{text}$ is the text-to-trope search function.

\subsubsection{Temperature}

To make the suggestions more diverse and less redundant, we introduce some randomness based on a final score accounting for their similarity score and their ranking to the model. We add a temperature parameter $\theta$ that controls the strength of the ranking over the distribution of the final scores, such as:
$$ s(T_i, T) = \left(\frac{N_T - rank_{(\tilde{s}, T_i)}(T)}{N_T}\right)^{\frac{1}{\theta}} * \tilde{s}(T_i, T) $$

where $rank_{(\tilde{s}, T_i)}(T)$ corresponds to the rank of the trope $T$ among all tropes based on their similarity $\tilde{s}$ to trope $T_i$, and $N_T$ is the total number of tropes considered.

Outputs are obtained by randomly drawing tropes without replacement. The probability of drawing a trope is proportional to its final score relative to the other final scores.

We empirically set $\theta$ to 0.02, deeming that it reasonably randomizes the outputs while providing satisfying suggestions following the score ranking.

\section{Suggestion evaluation} \label{sec:evaluation}

In this evaluation, we show that both of our methods provide valuable suggestions while having the intended characterizations.

\subsection{Methodology}

We conducted a within-subjects evaluation of our two algorithms, using as a baseline randomly generated tropes. As trope inputs, we randomly selected 36 tropes. For each input, we provided five trope propositions generated by each algorithm and the baseline. Participants were asked to rate six statements on a 7-point Likert scale about the set of propositions in relation to the initial input idea:
\begin{itemize}
    \item \textbf{S1-1:} I am familiar with the Initial Idea.
    \item \textbf{S1-2:} Each Proposition is often used with the Initial Idea.
    \item \textbf{S1-3:} Each Proposition shares similarities with the Initial Idea.
    \item \textbf{S1-4:} Each Proposition can be easily used with the Initial Idea.
    \item \textbf{S1-5:} Each Proposition offers a distinct direction to the story from the others.
    \item \textbf{S1-6:} I would use some of the Propositions to create my own story from the Initial Idea.
\end{itemize}
The first question allows us to verify that the input is understandable to the participant. Questions S1-2 to S1-5 enable us to characterize the outputs proposed by each algorithm. The last question determines if the algorithms suggestions are actually relevant to create a story.

Along with the first previous input trope, we then provided a randomly selected second input with a new list of five suggested tropes generated by one of the algorithms. Participants were asked to rate two more statements the same way:
\begin{itemize}
    \item \textbf{S2-1:} The Propositions combine the two Initial Ideas.
    \item \textbf{S2-2:} I would use some of the Propositions to create my own story from the two Initial Ideas.
\end{itemize}

Each participant rated nine distinct sets of inputs. Each algorithm generated suggestions for three of these input sets so that every participant rated each algorithm three times on different inputs. The order of the algorithm appearances was randomized. Each set of inputs handled by one of the algorithms was rated by at least five to 12 participants. Each trope was accompanied by a short description to help them understand the tropes. In total, we recruited 96 users on Prolific to participate in this evaluation. Participants were screened based on self-reporting enjoying and regularly engaging in creating stories.

\subsection{Results}

\begin{figure} 
\centering
\includegraphics[width=1\linewidth]{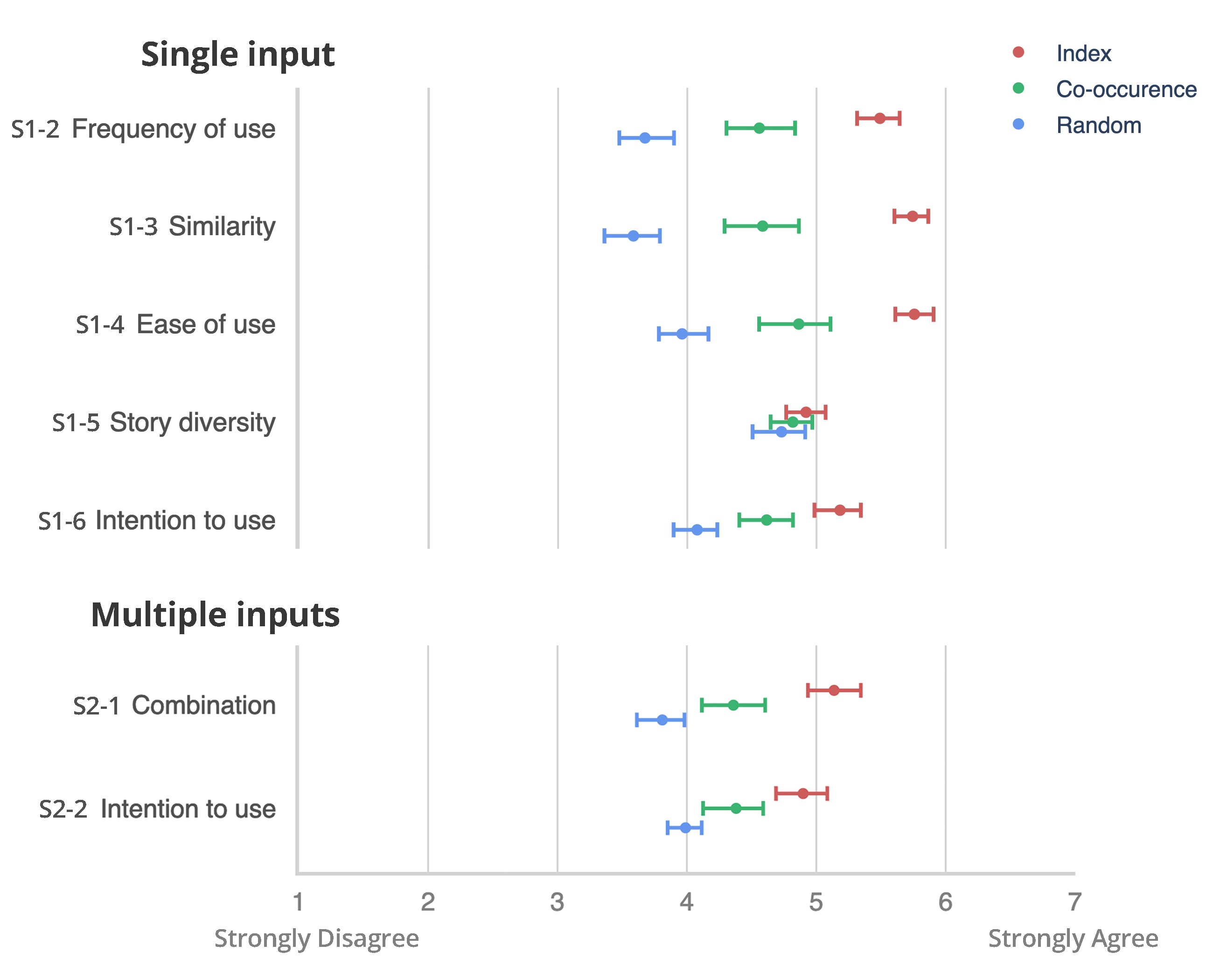}
\caption{Suggestion evaluation results for our proposed Index and Co-occurrence-based methods and a baseline generating random tropes.
}
\label{fig:results1}
\Description{Mean and error bars plot for each question displayed on the y-axis and 7-point Likert scale on the x-axis. Each algorithm is represented in a different color: red for the Index method, green for the Co-occurrence one, and blue for the Random one. The plot is divided into two parts. The top part corresponds to "Single Input" questions, the bottom part to "Multiple Inputs". For every question, it is clear that blue dots have the lowest 1-to-7 rate, green dots are in the middle, and red dots have the highest. S1-5 Story Diversity is the only question that is rated very closely, with the three dots and their error bars overlapping.}
\end{figure}

The results are displayed in Figure~\ref{fig:results1}. To analyze results, we only consider ratings where participants report being familiar with the initial input (at least 'Somewhat Agree' in S1-1). We removed 90 responses in which candidates declared being not familiar with the Initial Idea given as input for the evaluation. This represents 10\% of the total number of responses. We average the participants' ratings for each input and question to obtain a mean rating. With a mean standard deviation of 1.22 for the five to 12 answers, we consider that the participants agreed on the ratings. For each algorithm, we average the mean ratings obtained for each input and employ non-parametric bootstrapping \cite{efron_bootstrap_1979} with $R=1,000$ iterations to derive 95\% confidence intervals for all measures. 

For single inputs, the Index and Co-occurrence algorithms both provided suggestions that were more likely to be used than the baseline ($\mu = 5.18$, $\mu = 4.65$, and $\mu = 4.08$ respectively, for S1-6). For this question, the mean ratings were higher than the baseline for 97\% of the Index suggestions (35 out of 36 inputs) and for 75\% of the Co-occurrence suggestions (27 out of 36 inputs).  Although the Co-occurrence method relies on trope appearance with one another frequency, the Index method tropes were reported to be more frequently used with the input (Figure~\ref{fig:results1}). Overall, the Index algorithm was largely found to provide suggestions that were considered the most often used with (S1-2), similar to (S1-3), and easily usable with (S1-4) the input trope. Besides, we checked that the suggestions generated by the Index and Co-occurrence algorithms were almost fully distinct: on average, only 0.28 suggestion out of the five proposed were the same. As a result, we conclude that our algorithms provide different suggestions that would be rather useful for developing the input trope and that the Index algorithm provides suggestions that are most closely related to the input.

In addition, each of the three compared methods was similarly rated regarding to diversity ($\mu = 4.92$, $\mu = 4.82$, and $\mu = 4.73$ for Q1-5). In other words, our methods suggestions were deemed to provide story directions as distinct as random trope suggestions. Finally, we note that the "Frequency of use with," "Similarity," and "Ease of use" properties are strongly correlated when examining each participant's answer to each input ($p($S1-1$, $ S1-2$) = 0.80$, $p($S1-1$, $ S1-3$) = 0.68$, and $p($S1-2$, $ S1-3$) = 0.75$ for the random suggestions ratings), which may reflect some semantic overlap.

We observed similar results for multiple inputs in terms of intentions to use, with overall positive feedback for both our algorithms. However, the virtual adoption of the suggestions is this time lower, with middling Co-occurrence suggestions. The Index algorithm demonstrates better semantic combination of the two inputs ($\mu = 5.13$), compared to the Co-occurrence algorithm ($\mu = 4.36$) and the baseline ($\mu = 3.81$), showing the algorithms' difference in characterizations again.

\section{TaleStream Evaluation}\label{sec:study}

We conducted a user study to gain insights and feedback on the potential, limitations, and future opportunities of TaleStream and trope-based human-AI story co-creation. We focused on learning how our system and the use of tropes facilitate the exploration of story ideas and story co-creation. As this study seeks to understand the users' engagement with tropes and the system, we did not conduct a formal comparison to existing human-AI story co-creation tools but relied on participants' experiences.

\subsection{Participants}

We recruited 10 participants from our institutions by word-of-mouth and through mailing lists. Participants completed a screening survey about their writing background before the study. We selected 10 participants who reported writing stories at least a few times a week, including five hobbyists and five experts. We considered participants who reported writing stories for professional purposes as experts and those who wrote for personal enjoyment as hobbyists. Our participants came from diverse backgrounds, including journalism, theater, literature, improvisation, cinema, and role-playing games. Participants presented a wide variety of experiences providing a range of perspectives on the use of tropes in storytelling. Four experts had more than ten years of professional experience (U1, U2, U8, U10) in their field, and the fifth one had five (U9). We did not collect hobbyists’ years of experience. None of the participants had extensive experience with AI tools (the experiments were conducted before ChatGPT). Participants were compensated \$25 for the one-hour experiment. 

\begin{table}[h!]%3
    \centering
     \caption{Information about the participants of the system evaluation.}
     \label{table:tropekg1}  
    \begin{tabular}{ccc}
    \hline
    Participant & Experience & Fields\\
    \hline
    U1 & Expert & Animation, Film-making \\
    U2 & Expert & Advertisement, Film-making \\
    U3 & Hobbyist & Literature \\
    U4 & Hobbyist & Film-making \\
    U5 & Hobbyist & Improvisation, Theater \\
    U6 & Hobbyist & Literature \\
    U7 & Hobbyist & Role-play \\
    U8 & Expert & Film-making \\
    U9 & Expert & Theater \\
    U10 & Expert & Film-making \\
    \hline
    \end{tabular}
%    \Description{Three columns give participants' ID, experience, and fields.}
\end{table}

\subsection{Procedure}

We conducted remote studies on Zoom that were recorded. Participants were first given an overview of the study (5 min) before proceeding to a tutorial on using the tool (15 min). Participants used Chrome Remote Desktop to access the system on the interviewer’s computer. Since our goal was to understand how experienced story writers interacted with our system and tropes, we asked participants to use our system to imagine a new story by filling the canvas with elements that they wanted to use in their story. We specifically encouraged them to conceive a story that they had not thought about before. Participants were not expected to complete a full story and we did not restrict how or when they should use the suggestions. We asked participants to think aloud during the experiment to learn about their rationales and reactions in using the tool. Afterward, we conducted semi-structured interviews (20 min) about their experience with the system, asking them to reflect on their own practices to draw comparisons.

\begin{figure} 
\centering
\includegraphics[width=1\linewidth]{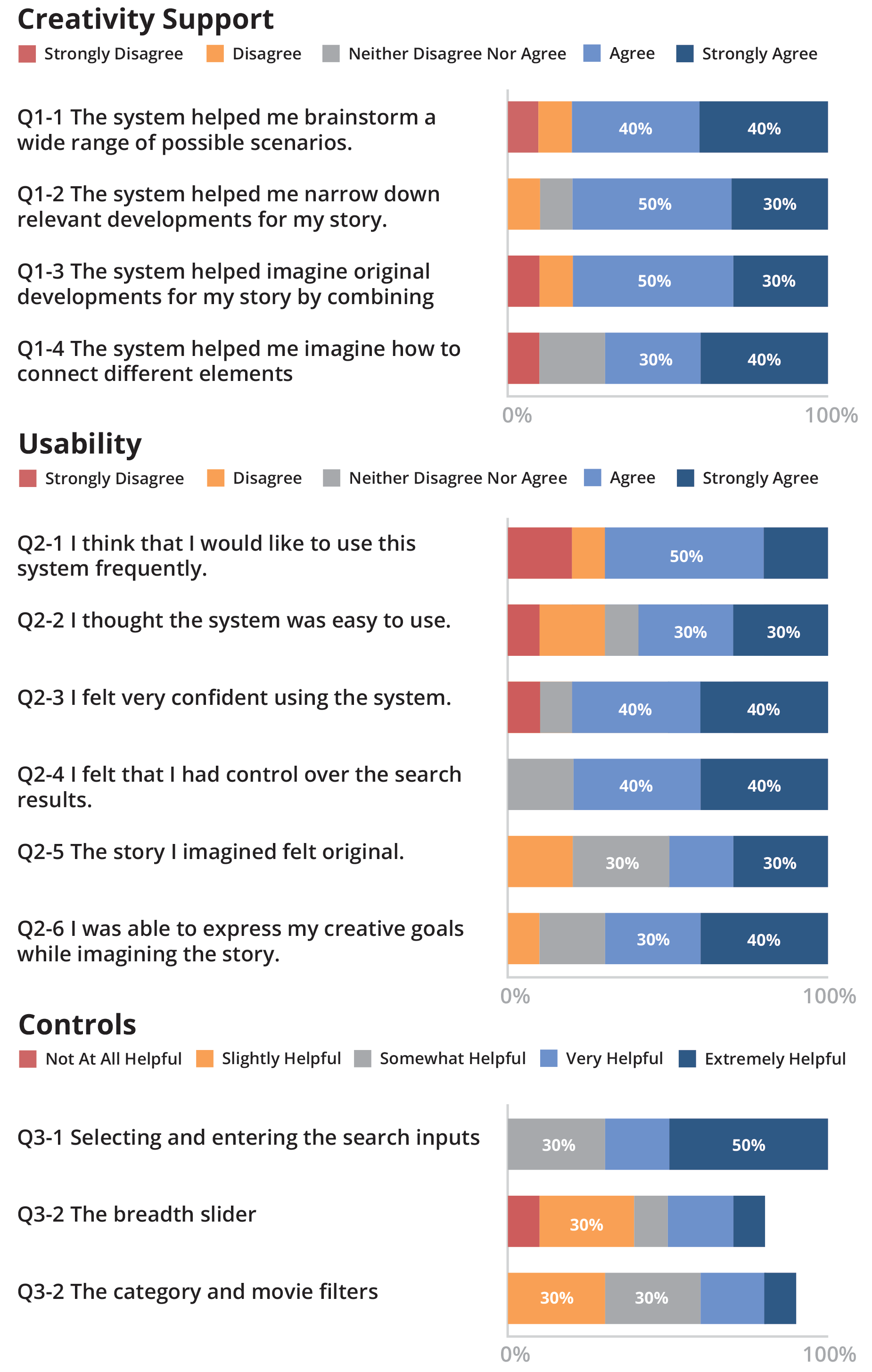}
\caption{Results of TaleStream evaluation survey on creativity support, usability, and controls.}
\label{fig:results2}
\Description{Horizontal bar plots with questions on the y-axis and the percentage of answers on the x-axis. The figure is divided into 3 parts with titles. From top to bottom: Creativity Support question, Usability questions, and Controls questions. Underneath the titles, legends of the bar colors are shown, going from red (Strongly Disagree / Not At All Helpful) to blue (Strongly Agree / Extremely Helpful).}
\end{figure}

Lastly, participants were asked to fill out a questionnaire divided into three sections. Our tool was designed as an aid to authors, we, therefore, focused on obtaining participants’ self-assessment of their results and their experience with the tool rather than relying on a third party’s assessment of their creations. In the first section, participants rated their level of agreement with questions related to the system's support for divergent and convergent thinking using a 5-point Likert scale. The second section assessed the system's usability, asking participants to rate their impressions on various aspects of the system's design and functionality. Finally, the third section focused on the helpfulness of the features controlling the suggestion results. Participants were asked to rate each feature on a scale from "Not at all helpful" to "Extremely helpful," with an additional option for "Did not use." The full list of questions can be found in Figure \ref{fig:results2}.

\section{Results}\label{sec:results}

\subsection{Creating with TaleStream}

\subsubsection{Creativity support}

In general, all participants expressed their enthusiasm for TaleStream: “It’s terrific. It was awesome. I love it.” (U4). 7 out of the 10 said they would use the system frequently (Q2-1 in Figure~\ref{fig:results2}). Reflecting on their creative processes and the tools they were familiar with, participants found TaleStream unique and complementary to their workflow.

All 10 participants deemed the system very useful and efficient for writer’s block, helping them to ideate and develop compelling stories in a few minutes, which they would normally struggle to do in days in regular workflows. TaleStream supported convergent thinking. 8 participants agreed that the system helped them narrow down the possibilities and focus on some ideas (Q1-2). U10 liked that some ideas were “far more specific”, which “helped [them] specify and decide on which tropes [they] actually wanted to pursue”. Besides, the system helped connect different ideas (Q1-4) for 7 participants. U1 saw the system as “a GPS that takes [them] from point A to point B without replacing A or B”, helping them figure out how to fill out some holes.

The system also supported divergent thinking. The system expanded the participants’ range of story possibilities according to 8 participants (Q1-1). For U4, “inputting a term or word [branched] out almost like a spider web and [gave] different options”. Finally, the system’s suggestions combining ideas helped 8 participants to broaden their insights (Q1-3). For instance, U10 was surprised to be able to find tropes combining concepts related to Western and Sci-Fi and was inspired by the “mashup of different types of tropes” they obtained.

\subsubsection{Collaborating with TaleStream}

Most participants (8/10) felt they had adequate controllability over the suggestion results  (Figure~\ref{fig:results2} Q2-4). Participants found selecting the inputs and entering text particularly helpful in guiding the search (Q3-1). Participants showed diverse opinions about the breadth slider and the filters (Q3-2 and Q3-3). In open-ended feedback, some participants described forgetting about or leaving aside the breadth feature due to the lack of time. The filters were used by half of the participants specifically when looking for inspiration for specific narrative holes, such as characters and settings (U6, U7), or from particular movies (U7, U10) for instance. Overall, participants found the controls understandable and easily usable to steer the suggestion results.

The system’s effectiveness, freedom (Q2-6), and ease of use (Q2-2) helped build confidence (Q2-3) and trust with the participants who felt they were collaborating with the system, referring to it as “a storyteller assistant” (U1), an “effective brainstorming partner” (U6), or “a writing partner” (U10).

\subsection{Using tropes}

\subsubsection{A common story lexicon}

While 6 out of the 10 participants had not previously heard about the concept of trope, all participants were familiar with most of the encountered tropes. Some even reckoned using these story mechanisms all the time, mostly unknowingly (U1, U5, U7, U9). This natural familiarity makes tropes particularly evocative, allowing participants to get a quick picture of possible stories (U1, U3, U5, U7, U10): “It gave me a visual of what could be happening much faster than having to write it down” (U3). This common language was also deemed to be excellent for quickly communicating ideas (U8, U9) and for “getting everyone on the same page” (U8), which are key goals when making canvases.

This common language is built on a history of occurrences that participants enjoyed having access to easily. It helped them better understand some tropes (U7, U8, U10) and explore implementations of tropes (U4, U5, U7, U8, U9). Exploring the movies \textit{The Groundhog Day} and \textit{Looper}’s tropes, U7 stated: “The most important thing for me is understanding what this particular character or structure is, and then being able to translate that to a different set”. However, U2 and U6 felt a bit overwhelmed by the number of examples they were unfamiliar with, wishing they could filter some. Besides, U6 was “afraid of looking at the examples” because they did not want to over-rely on them.

\subsubsection{Reinventing tropes}

Originality is indeed often a concern when using tropes. Many participants mentioned overreliance on tropes as a trap (U2, U5, U6, U7, U8). All participants reported that TaleStream pushed them to be more conscious of these misuses, making U1 “more confident in building the story” or helping U4 “not plagiarize because the idea is out there”. Once aware of these pitfalls, tropes are great assets for story creation. All participants pointed out the diversity of uses of a single trope and the opportunity to reinvent each of them. With TaleStream, tropes were never seen as rigid building blocks, but as loose ideas that can be freely interpreted, interrogating more than imposing. As such, and despite relying on tropes to build their stories, participants generally considered their story created in less than 20 minutes original (Q3-4).

\subsubsection{Building blocks}

Within TaleStream, tropes were used as building blocks of stories. On average, more than half of the final canvas elements were tropes. Most of these tropes were suggested by TaleStream based on the participants' inputs. Additionally, half of the text elements were directly used in conjunction with a specific trope to develop or detail it. U8 compared tropes to a “pre-built Lego set”, while U2 believes that “stories are made of other stories”. Overall, creating with tropes helped participants envision their stories at a high level, making them more aware of the underlying structure and increasing their confidence in the creation process. Participants used tropes to start building the “scaffolding” (U7) of their stories (U1, U5, U7, U10). Tropes then served to flesh out the backbone (U1, U2, U3, U5, U8, U9, U10) with “smaller and smaller details, almost like brushstrokes” (U3). Finally, participants described how they implemented the tropes in their stories in their own words, which felt like “coloring a story” (U5).

\subsection{Summary}

Overall, our study suggests that TaleStream provides an original and effective way to build stories with controls that support writers in their creative flow. Tropes, as story building blocks, proved to be a shared language among the participants, which they naturally adopted to explore existing works as well as imagine and conceive their stories. Working with tropes helped participants be more confident in the creation process by making them aware of the underlying structure and patterns that constitute the core of their stories.

\section{Limitations and Future Work} \label{sec:lim_future}
\subsection{Experiments limitations}

Both of our experiments present limitations to consider. The mixed algorithm slider was introduced to the system only after conducting our suggestion evaluation of the Index and Co-occurrence algorithms for trope suggestions which revealed that both algorithms produced useful but distinct trope suggestions. Although we did not directly evaluate the mixed algorithm, participants in our system study did use the mixed slider to adjust the suggestions according to their needs and reported their satisfaction in both the questionnaire and the interview.

In the evaluation of TaleStream, our objective was to analyze the interactions between story writers and the system. However, the experiments did not entirely represent how a writer would use our tool. The study was limited to a 20-minute story ideation task, which may not have allowed participants to fully grasp the system. Participants were also asked to generate original ideas from scratch, which occasionally felt artificial (U6, U9). All participants expressed curiosity in trying TaleStream throughout the entire story creation process, not just at the beginning. It is also important to note that the insights derived from our results may not capture the characteristics of all workflows and that the features and design of our system may not be optimal for all cases. Our participants represented a relatively diverse range of backgrounds, each requiring different modes of thinking and creation support. Additionally, storytelling expertise could also influence the user experience. Although we did not observe significant differences in the interactions with the system between experts and hobbyists, our experiment was not designed to point them out. Our only observation was that expert story writers were more familiar with tropes and reported commonly using tropes (U1, U2) or having learned about them during their curriculum (U10). To obtain more realistic results, a randomized controlled and longitudinal field trial involving story writers over an extended period would be beneficial in exploring how creators perceive and integrate TaleStream into their existing creative workflows in different contexts.

\subsection{Developing the trope framework}

In this work, tropes were employed as a framework to approach narrative design, serving as building blocks. Our evaluation results highlighted several promising directions for further development. Overall, tropes prove to be effective elements for navigating the story space and obtaining tailored suggestions. We could explore additional mechanisms to support the search for tropes. For instance, incorporating category filter suggestions could assist users in refining their searches. We could also leverage insights from systems that incorporate users' direct feedback or analyze their activity \cite{koch_may_2019} and develop analogous suggestions adaptation mechanisms for tropes.

Aside from tropes, canvases typically rely on visual elements. Many participants expressed their desire for visual aids to complement the tropes, which would be beneficial for grounding ideas (U2, U8, U9, U10). To illustrate specific tropes, images could either be directly extracted from scenes in which they appear or generated based on wiki data. Other participants desired more detailed suggestions, such as "random details" about what characters have for breakfast within the Morning Routine trope (U7). This level of specificity could be provided through fine-tuned text generation using trope examples from tvtropes.org. More broadly, tropes could be used as high-level controls for generating text and guiding the flow of the story. Some participants envisioned a side text editor linked to the canvas and its elements. Within this framework, it could be easier and more natural for humans to express their creative intentions to computers while enabling computers to respond according to the story structure and stakes.

\subsection{Inclusiveness and awareness}

Tropes can perpetuate a narrow perspective, harmful representations, and stereotypes that can negatively impact individuals and communities. While tropes serve as convenient shortcuts for conveying familiar ideas, they can also hinder the exploration of more nuanced narratives by constraining people's imagination and inadvertently promoting laziness. Other biases can strengthen this danger. Our dataset was extracted from the English version of tvtropes.org, which strongly focuses on popular Western culture: U9 could not find a specific Kenyan film in the filters. It is essential for creators and consumers to be critical of these tropes, actively challenging and dismantling them to foster a more inclusive and equitable cultural landscape. One important challenge for future work is to address the limitations and biases of the dataset that will almost inexorably lead to biased results \cite{mitchell_model_2019}. By aiming for greater diversity, inclusivity, and comprehensiveness in media representation, we can foster more nuanced storytelling systems. This involves seeking annotations from a broader array of sources, including non-Western references, to ensure a richer and more culturally varied narrative landscape. This could be accomplished by incorporating data from tvtropes.org in different languages or by developing automated methods for trope detection, an active area of research \cite{su_truman_2021, chang_situation_2021}. Several other potential approaches can be considered to address biases. Firstly, implementing mechanisms to flag or censor tropes that are deemed problematic can help mitigate biases. Additionally, providing contextual guidelines that encourage critical thinking or offering examples of how to subvert each suggested trope\footnote[1]{Following the \href{https://tvtropes.org/pmwiki/pmwiki.php/Main/PlayingWithATrope}{\textit{Playing with a Trope}} article from tvtropes.org} to propose alternative viewpoints are strategies to explore.

\section{Conclusion} \label{sec:conc}

In this paper, we introduce TaleStream, a canvas system that suggests story ideas in the form of tropes. The trope and text elements on the canvas can be selected to generate trope suggestions which can be explored with movie examples and steered with additional controls on the story space. Our technical evaluation of the suggestion algorithms shows that our methods provide valuable results with different characterizations. The system evaluation revealed that TaleStream supports creative abilities for story ideation, provides reliable controllability, and is perceived as a pleasant partner accompanying the creative flow. The use of tropes in TaleStream was found to be particularly effective for quickly visualizing ideas through references, being aware of common pitfalls, and structuring stories, making users more confident while creating. This work opens up new ways to leverage tropes to support story creation as an intermediate comprehensive lexicon of storytelling.

\begin{acks}
We thank Charly Lehuédé, Egan Tizzoni, Loïc Matos, Marc Löning, Ian-Christopher Tanoh, and Saehui Hwang for  valuable conversations, our user study participants for their insights, and the reviewers for their feedback. The first author was partially supported by the Brown Institute for Media and Innovation.
\end{acks}

\balance
\bibliographystyle{ACM-Reference-Format}
\bibliography{paper}


\begin{thebibliography}{76}

%%% the Web address.

\ifx \showCODEN    \undefined \def \showCODEN     #1{\unskip}     \fi
\ifx \showDOI      \undefined \def \showDOI       #1{#1}\fi
\ifx \showISBNx    \undefined \def \showISBNx     #1{\unskip}     \fi
\ifx \showISBNxiii \undefined \def \showISBNxiii  #1{\unskip}     \fi
\ifx \showISSN     \undefined \def \showISSN      #1{\unskip}     \fi
\ifx \showLCCN     \undefined \def \showLCCN      #1{\unskip}     \fi
\ifx \shownote     \undefined \def \shownote      #1{#1}          \fi
\ifx \showarticletitle \undefined \def \showarticletitle #1{#1}   \fi
\ifx \showURL      \undefined \def \showURL       {\relax}        \fi
% invisible to TeX
\providecommand\bibfield[2]{#2}
\providecommand\bibinfo[2]{#2}
\providecommand\natexlab[1]{#1}
\providecommand\showeprint[2][]{arXiv:#2}

\bibitem[\protect\citeauthoryear{??}{noa}{1994}]%
        {noauthor_dramatica_nodate}
 \bibinfo{year}{1994}\natexlab{}.
\newblock \bibinfo{title}{Dramatica}.
\newblock
\newblock
\urldef\tempurl%
\url{https://dramatica.com/}
\showURL{%
\tempurl}


\bibitem[\protect\citeauthoryear{??}{noa}{2019}]%
        {noauthor_ai_nodate}
 \bibinfo{year}{2019}\natexlab{}.
\newblock \bibinfo{title}{{AI} {Dungeon}}.
\newblock
\newblock
\urldef\tempurl%
\url{https://play.aidungeon.io/main/home}
\showURL{%
\tempurl}


\bibitem[\protect\citeauthoryear{??}{noa}{2020}]%
        {noauthor_plottr_nodate}
 \bibinfo{year}{2020}\natexlab{}.
\newblock \bibinfo{title}{Plottr}.
\newblock
\newblock
\urldef\tempurl%
\url{https://plottr.com/}
\showURL{%
\tempurl}


\bibitem[\protect\citeauthoryear{??}{noa}{2021}]%
        {noauthor_sudowrite_nodate}
 \bibinfo{year}{2021}\natexlab{}.
\newblock \bibinfo{title}{Sudowrite}.
\newblock
\newblock
\urldef\tempurl%
\url{https://www.sudowrite.com/}
\showURL{%
\tempurl}


\bibitem[\protect\citeauthoryear{??}{noa}{2022}]%
        {noauthor_introducing_nodate}
 \bibinfo{year}{2022}\natexlab{}.
\newblock \bibinfo{title}{Introducing {ChatGPT}}.
\newblock
\newblock
\urldef\tempurl%
\url{https://openai.com/blog/chatgpt}
\showURL{%
\tempurl}


\bibitem[\protect\citeauthoryear{Akoury, Wang, Whiting, Hood, Peng, and
  Iyyer}{Akoury et~al\mbox{.}}{2020}]%
        {akoury_storium_2020}
\bibfield{author}{\bibinfo{person}{Nader Akoury}, \bibinfo{person}{Shufan
  Wang}, \bibinfo{person}{Josh Whiting}, \bibinfo{person}{Stephen Hood},
  \bibinfo{person}{Nanyun Peng}, {and} \bibinfo{person}{Mohit Iyyer}.}
  \bibinfo{year}{2020}\natexlab{}.
\newblock \showarticletitle{{STORIUM}: {A} {Dataset} and evaluation platform
  for machine-in-the-loop story generation}.
\newblock \bibinfo{journal}{\emph{arXiv preprint arXiv:2010.01717}}
  (\bibinfo{year}{2020}).
\newblock


\bibitem[\protect\citeauthoryear{Alhussain and Azmi}{Alhussain and
  Azmi}{2021}]%
        {alhussain_automatic_2021}
\bibfield{author}{\bibinfo{person}{Arwa~I. Alhussain} {and}
  \bibinfo{person}{Aqil~M. Azmi}.} \bibinfo{year}{2021}\natexlab{}.
\newblock \showarticletitle{Automatic {Story} {Generation}: {A} {Survey} of
  {Approaches}}.
\newblock \bibinfo{journal}{\emph{Comput. Surveys}} \bibinfo{volume}{54},
  \bibinfo{number}{5} (\bibinfo{date}{May} \bibinfo{year}{2021}),
  \bibinfo{pages}{103:1--103:38}.
\newblock
\showISSN{0360-0300}
\urldef\tempurl%
\url{https://doi.org/10.1145/3453156}
\showDOI{\tempurl}


\bibitem[\protect\citeauthoryear{Alvarez and Font}{Alvarez and Font}{2022}]%
        {alvarez_tropetwist_2022}
\bibfield{author}{\bibinfo{person}{Alberto Alvarez} {and} \bibinfo{person}{Jose
  Font}.} \bibinfo{year}{2022}\natexlab{}.
\newblock \showarticletitle{{TropeTwist}: {Trope}-{Based} {Narrative}
  {Structure} {Generation}}. In \bibinfo{booktitle}{\emph{Proceedings of the
  17th {International} {Conference} on the {Foundations} of {Digital} {Games}}}
  \emph{(\bibinfo{series}{{FDG} '22})}. \bibinfo{publisher}{Association for
  Computing Machinery}, \bibinfo{address}{New York, NY, USA}.
\newblock
\showISBNx{978-1-4503-9795-7}
\urldef\tempurl%
\url{https://doi.org/10.1145/3555858.3563271}
\showDOI{\tempurl}
\newblock
\shownote{event-place: Athens, Greece.}


\bibitem[\protect\citeauthoryear{Alvarez, Font, and Togelius}{Alvarez
  et~al\mbox{.}}{2022}]%
        {alvarez_story_2022}
\bibfield{author}{\bibinfo{person}{Alberto Alvarez}, \bibinfo{person}{Jose
  Font}, {and} \bibinfo{person}{Julian Togelius}.}
  \bibinfo{year}{2022}\natexlab{}.
\newblock \showarticletitle{Story {Designer}: {Towards} a {Mixed}-{Initiative}
  {Tool} to {Create} {Narrative} {Structures}}. In
  \bibinfo{booktitle}{\emph{Proceedings of the 17th {International}
  {Conference} on the {Foundations} of {Digital} {Games}}}
  \emph{(\bibinfo{series}{{FDG} '22})}. \bibinfo{publisher}{Association for
  Computing Machinery}, \bibinfo{address}{New York, NY, USA}.
\newblock
\showISBNx{978-1-4503-9795-7}
\urldef\tempurl%
\url{https://doi.org/10.1145/3555858.3555929}
\showDOI{\tempurl}
\newblock
\shownote{event-place: Athens, Greece.}


\bibitem[\protect\citeauthoryear{Ammanabrolu, Cheung, Broniec, and
  Riedl}{Ammanabrolu et~al\mbox{.}}{2020}]%
        {ammanabrolu_automated_2020}
\bibfield{author}{\bibinfo{person}{Prithviraj Ammanabrolu},
  \bibinfo{person}{Wesley Cheung}, \bibinfo{person}{William Broniec}, {and}
  \bibinfo{person}{Mark~O. Riedl}.} \bibinfo{year}{2020}\natexlab{}.
\newblock \bibinfo{title}{Automated {Storytelling} via {Causal}, {Commonsense}
  {Plot} {Ordering}}.
\newblock
\newblock
\urldef\tempurl%
\url{https://doi.org/10.48550/arXiv.2009.00829}
\showDOI{\tempurl}
\newblock
\shownote{arXiv:2009.00829 [cs].}


\bibitem[\protect\citeauthoryear{Aristotle}{Aristotle}{2006}]%
        {aristotle_poetics_2006}
\bibfield{author}{\bibinfo{person}{Aristotle}.}
  \bibinfo{year}{2006}\natexlab{}.
\newblock \bibinfo{booktitle}{\emph{Poetics}}.
\newblock \bibinfo{publisher}{ReadHowYouWant.com}.
\newblock
\showISBNx{978-1-4250-0095-0}
\newblock
\shownote{Google-Books-ID: sywjT24pBb8C.}


\bibitem[\protect\citeauthoryear{Bamman, O'Connor, and Smith}{Bamman
  et~al\mbox{.}}{2013}]%
        {bamman_learning_2013}
\bibfield{author}{\bibinfo{person}{David Bamman}, \bibinfo{person}{Brendan~T.
  O'Connor}, {and} \bibinfo{person}{Noah~A. Smith}.}
  \bibinfo{year}{2013}\natexlab{}.
\newblock \showarticletitle{Learning {Latent} {Personas} of {Film}
  {Characters}}. In \bibinfo{booktitle}{\emph{Annual {Meeting} of the
  {Association} for {Computational} {Linguistics}}}.
\newblock


\bibitem[\protect\citeauthoryear{Barraza-Urbina}{Barraza-Urbina}{2017}]%
        {barraza-urbina_exploration-exploitation_2017}
\bibfield{author}{\bibinfo{person}{Andrea Barraza-Urbina}.}
  \bibinfo{year}{2017}\natexlab{}.
\newblock \showarticletitle{The {Exploration}-{Exploitation} {Trade}-off in
  {Interactive} {Recommender} {Systems}}. In
  \bibinfo{booktitle}{\emph{Proceedings of the {Eleventh} {ACM} {Conference} on
  {Recommender} {Systems}}} \emph{(\bibinfo{series}{{RecSys} '17})}.
  \bibinfo{publisher}{Association for Computing Machinery},
  \bibinfo{address}{New York, NY, USA}, \bibinfo{pages}{431--435}.
\newblock
\showISBNx{978-1-4503-4652-8}
\urldef\tempurl%
\url{https://doi.org/10.1145/3109859.3109866}
\showDOI{\tempurl}


\bibitem[\protect\citeauthoryear{Biermann, Ma, and Yoon}{Biermann
  et~al\mbox{.}}{2022}]%
        {biermann_from_2022}
\bibfield{author}{\bibinfo{person}{Oloff~C. Biermann}, \bibinfo{person}{Ning~F.
  Ma}, {and} \bibinfo{person}{Dongwook Yoon}.} \bibinfo{year}{2022}\natexlab{}.
\newblock \showarticletitle{From Tool to Companion: Storywriters Want AI
  Writers to Respect Their Personal Values and Writing Strategies}. In
  \bibinfo{booktitle}{\emph{Proceedings of the 2022 ACM Designing Interactive
  Systems Conference}} (Virtual Event, Australia) \emph{(\bibinfo{series}{DIS
  '22})}. \bibinfo{publisher}{Association for Computing Machinery},
  \bibinfo{address}{New York, NY, USA}, \bibinfo{pages}{1209–1227}.
\newblock
\showISBNx{9781450393584}
\urldef\tempurl%
\url{https://doi.org/10.1145/3532106.3533506}
\showDOI{\tempurl}


\bibitem[\protect\citeauthoryear{Boden and Boden}{Boden and Boden}{2004}]%
        {boden_creative_2004}
\bibfield{author}{\bibinfo{person}{Margaret~A. Boden} {and}
  \bibinfo{person}{Research Professor of Cognitive Science Margaret~A. Boden}.}
  \bibinfo{year}{2004}\natexlab{}.
\newblock \bibinfo{booktitle}{\emph{The {Creative} {Mind}: {Myths} and
  {Mechanisms}}}.
\newblock \bibinfo{publisher}{Psychology Press}.
\newblock
\showISBNx{978-0-415-31452-7}
\newblock
\shownote{Google-Books-ID: 6Zkm4dz32Y4C.}


\bibitem[\protect\citeauthoryear{Booker}{Booker}{2004}]%
        {booker_seven_2004}
\bibfield{author}{\bibinfo{person}{Christopher Booker}.}
  \bibinfo{year}{2004}\natexlab{}.
\newblock \bibinfo{booktitle}{\emph{The {Seven} {Basic} {Plots}: {Why} {We}
  {Tell} {Stories}}}.
\newblock \bibinfo{publisher}{A\&C Black}.
\newblock
\showISBNx{978-0-8264-8037-8}
\newblock
\shownote{Google-Books-ID: XEUamcjBo9IC.}


\bibitem[\protect\citeauthoryear{Calderwood, Qiu, Gero, and Chilton}{Calderwood
  et~al\mbox{.}}{2020}]%
        {calderwood_how_2020}
\bibfield{author}{\bibinfo{person}{Alex Calderwood}, \bibinfo{person}{Vivian
  Qiu}, \bibinfo{person}{K. Gero}, {and} \bibinfo{person}{Lydia~B. Chilton}.}
  \bibinfo{year}{2020}\natexlab{}.
\newblock \showarticletitle{How {Novelists} {Use} {Generative} {Language}
  {Models}: {An} {Exploratory} {User} {Study}}.
\newblock
\urldef\tempurl%
\url{https://www.semanticscholar.org/paper/How-Novelists-Use-Generative-Language-Models%3A-An-Calderwood-Qiu/8cf1fc0b87dfda2a11bfaaaa3a0bf9f9e069bb0f}
\showURL{%
\tempurl}


\bibitem[\protect\citeauthoryear{Campbell}{Campbell}{2008}]%
        {campbell_hero_2008}
\bibfield{author}{\bibinfo{person}{Joseph Campbell}.}
  \bibinfo{year}{2008}\natexlab{}.
\newblock \bibinfo{booktitle}{\emph{The {Hero} with a {Thousand} {Faces}}}.
\newblock \bibinfo{publisher}{New World Library}.
\newblock
\showISBNx{978-1-57731-593-3}
\newblock
\shownote{Google-Books-ID: I1uFuXlvFgMC.}


\bibitem[\protect\citeauthoryear{Castricato, Frazier, Balloch, and
  Riedl}{Castricato et~al\mbox{.}}{2021}]%
        {castricato_fabula_2021}
\bibfield{author}{\bibinfo{person}{Louis Castricato}, \bibinfo{person}{Spencer
  Frazier}, \bibinfo{person}{Jonathan Balloch}, {and} \bibinfo{person}{Mark
  Riedl}.} \bibinfo{year}{2021}\natexlab{}.
\newblock \showarticletitle{Fabula {Entropy} {Indexing}: {Objective} {Measures}
  of {Story} {Coherence}}. In \bibinfo{booktitle}{\emph{Proceedings of the
  {Third} {Workshop} on {Narrative} {Understanding}}}.
  \bibinfo{publisher}{Association for Computational Linguistics},
  \bibinfo{address}{Virtual}, \bibinfo{pages}{84--94}.
\newblock
\urldef\tempurl%
\url{https://doi.org/10.18653/v1/2021.nuse-1.9}
\showDOI{\tempurl}


\bibitem[\protect\citeauthoryear{Cavazza, Charles, and Mead}{Cavazza
  et~al\mbox{.}}{2001}]%
        {cavazza_characters_2001}
\bibfield{author}{\bibinfo{person}{Marc Cavazza}, \bibinfo{person}{Fred
  Charles}, {and} \bibinfo{person}{Steven~J. Mead}.}
  \bibinfo{year}{2001}\natexlab{}.
\newblock \showarticletitle{Characters in {Search} of an {Author}: {AI}-{Based}
  {Virtual} {Storytelling}}. In \bibinfo{booktitle}{\emph{Virtual
  {Storytelling} {Using} {Virtual} {Reality} {Technologies} for
  {Storytelling}}} \emph{(\bibinfo{series}{Lecture {Notes} in {Computer}
  {Science}})}, \bibfield{editor}{\bibinfo{person}{Olivier Balet},
  \bibinfo{person}{Gérard Subsol}, {and} \bibinfo{person}{Patrice Torguet}}
  (Eds.). \bibinfo{publisher}{Springer}, \bibinfo{address}{Berlin, Heidelberg},
  \bibinfo{pages}{145--154}.
\newblock
\showISBNx{978-3-540-45420-5}
\urldef\tempurl%
\url{https://doi.org/10.1007/3-540-45420-9_16}
\showDOI{\tempurl}


\bibitem[\protect\citeauthoryear{Chang, Su, Hsu, Wang, Chang, Liu, Chang,
  Cheng, Wang, and Hsu}{Chang et~al\mbox{.}}{2021}]%
        {chang_situation_2021}
\bibfield{author}{\bibinfo{person}{Chen-Hsi Chang}, \bibinfo{person}{Hung-Ting
  Su}, \bibinfo{person}{Jui-Heng Hsu}, \bibinfo{person}{Yu-Siang Wang},
  \bibinfo{person}{Yu-Cheng Chang}, \bibinfo{person}{Zhe~Yu Liu},
  \bibinfo{person}{Ya-Liang Chang}, \bibinfo{person}{Wen-Feng Cheng},
  \bibinfo{person}{Ke-Jyun Wang}, {and} \bibinfo{person}{Winston~H. Hsu}.}
  \bibinfo{year}{2021}\natexlab{}.
\newblock \showarticletitle{Situation and {Behavior} {Understanding} by {Trope}
  {Detection} on {Films}}. In \bibinfo{booktitle}{\emph{Proceedings of the
  {Web} {Conference} 2021}} \emph{(\bibinfo{series}{{WWW} '21})}.
  \bibinfo{publisher}{Association for Computing Machinery},
  \bibinfo{address}{New York, NY, USA}, \bibinfo{pages}{3188--3198}.
\newblock
\showISBNx{978-1-4503-8312-7}
\urldef\tempurl%
\url{https://doi.org/10.1145/3442381.3449806}
\showDOI{\tempurl}


\bibitem[\protect\citeauthoryear{Chou and Christie}{Chou and Christie}{2021}]%
        {chou_structures_2021}
\bibfield{author}{\bibinfo{person}{Jean-Peic Chou} {and} \bibinfo{person}{Marc
  Christie}.} \bibinfo{year}{2021}\natexlab{}.
\newblock \showarticletitle{Structures in {Tropes} {Networks}: {Toward} a
  {Formal} {Story} {Grammar}}. In \bibinfo{booktitle}{\emph{Proceedings of the
  {Twelfth} {International} {Conference} on {Computational} {Creativity}}}.
  \bibinfo{address}{Mexico, Mexico}.
\newblock
\urldef\tempurl%
\url{https://hal.inria.fr/hal-03777738}
\showURL{%
\tempurl}


\bibitem[\protect\citeauthoryear{Chung, Kim, Yoo, Lee, Adar, and Chang}{Chung
  et~al\mbox{.}}{2022}]%
        {chung_talebrush_2022}
\bibfield{author}{\bibinfo{person}{John Joon~Young Chung},
  \bibinfo{person}{Wooseok Kim}, \bibinfo{person}{Kang~Min Yoo},
  \bibinfo{person}{Hwaran Lee}, \bibinfo{person}{Eytan Adar}, {and}
  \bibinfo{person}{Minsuk Chang}.} \bibinfo{year}{2022}\natexlab{}.
\newblock \showarticletitle{{TaleBrush}: {Sketching} {Stories} with
  {Generative} {Pretrained} {Language} {Models}}. In
  \bibinfo{booktitle}{\emph{Proceedings of the 2022 {CHI} {Conference} on
  {Human} {Factors} in {Computing} {Systems}}} \emph{(\bibinfo{series}{{CHI}
  '22})}. \bibinfo{publisher}{Association for Computing Machinery},
  \bibinfo{address}{New York, NY, USA}.
\newblock
\showISBNx{978-1-4503-9157-3}
\urldef\tempurl%
\url{https://doi.org/10.1145/3491102.3501819}
\showDOI{\tempurl}
\newblock
\shownote{event-place: New Orleans, LA, USA.}


\bibitem[\protect\citeauthoryear{Clark, Ross, Tan, Ji, and Smith}{Clark
  et~al\mbox{.}}{2018}]%
        {clark_creative_2018}
\bibfield{author}{\bibinfo{person}{Elizabeth Clark},
  \bibinfo{person}{Anne~Spencer Ross}, \bibinfo{person}{Chenhao Tan},
  \bibinfo{person}{Yangfeng Ji}, {and} \bibinfo{person}{Noah~A. Smith}.}
  \bibinfo{year}{2018}\natexlab{}.
\newblock \showarticletitle{Creative {Writing} with a {Machine} in the {Loop}:
  {Case} {Studies} on {Slogans} and {Stories}}. In
  \bibinfo{booktitle}{\emph{23rd {International} {Conference} on {Intelligent}
  {User} {Interfaces}}} \emph{(\bibinfo{series}{{IUI} '18})}.
  \bibinfo{publisher}{Association for Computing Machinery},
  \bibinfo{address}{New York, NY, USA}, \bibinfo{pages}{329--340}.
\newblock
\showISBNx{978-1-4503-4945-1}
\urldef\tempurl%
\url{https://doi.org/10.1145/3172944.3172983}
\showDOI{\tempurl}


\bibitem[\protect\citeauthoryear{Cook}{Cook}{2011}]%
        {cook_plotto_2011}
\bibfield{author}{\bibinfo{person}{William Cook}.}
  \bibinfo{year}{2011}\natexlab{}.
\newblock \bibinfo{booktitle}{\emph{{PLOTTO}: the master book of all plots}}.
\newblock \bibinfo{publisher}{Tin House Books}.
\newblock


\bibitem[\protect\citeauthoryear{Datta, Kovaleva, Mardziel, and Sen}{Datta
  et~al\mbox{.}}{2017}]%
        {datta_latent_2017}
\bibfield{author}{\bibinfo{person}{Anupam Datta}, \bibinfo{person}{Sophia
  Kovaleva}, \bibinfo{person}{Piotr Mardziel}, {and} \bibinfo{person}{Shayak
  Sen}.} \bibinfo{year}{2017}\natexlab{}.
\newblock \showarticletitle{Latent factor interpretations for collaborative
  filtering}.
\newblock \bibinfo{journal}{\emph{arXiv preprint arXiv:1711.10816}}
  (\bibinfo{year}{2017}).
\newblock


\bibitem[\protect\citeauthoryear{Duval, Lamson, de~Léséleuc~de Kérouara, and
  Gallé}{Duval et~al\mbox{.}}{2021}]%
        {duval_breaking_2021}
\bibfield{author}{\bibinfo{person}{Alexandre Duval}, \bibinfo{person}{Thomas
  Lamson}, \bibinfo{person}{Gaël de~Léséleuc~de Kérouara}, {and}
  \bibinfo{person}{Matthias Gallé}.} \bibinfo{year}{2021}\natexlab{}.
\newblock \showarticletitle{Breaking {Writer}'s {Block}: {Low}-cost
  {Fine}-tuning of {Natural} {Language} {Generation} {Models}}. In
  \bibinfo{booktitle}{\emph{Proceedings of the 16th {Conference} of the
  {European} {Chapter} of the {Association} for {Computational} {Linguistics}:
  {System} {Demonstrations}}}. \bibinfo{publisher}{Association for
  Computational Linguistics}, \bibinfo{address}{Online},
  \bibinfo{pages}{278--287}.
\newblock
\urldef\tempurl%
\url{https://doi.org/10.18653/v1/2021.eacl-demos.33}
\showDOI{\tempurl}


\bibitem[\protect\citeauthoryear{Efron}{Efron}{1979}]%
        {efron_bootstrap_1979}
\bibfield{author}{\bibinfo{person}{B. Efron}.} \bibinfo{year}{1979}\natexlab{}.
\newblock \showarticletitle{Bootstrap {Methods}: {Another} {Look} at the
  {Jackknife}}.
\newblock \bibinfo{journal}{\emph{The Annals of Statistics}}
  \bibinfo{volume}{7}, \bibinfo{number}{1} (\bibinfo{year}{1979}),
  \bibinfo{pages}{1--26}.
\newblock
\showISSN{0090-5364}
\urldef\tempurl%
\url{https://www.jstor.org/stable/2958830}
\showURL{%
\tempurl}
\newblock
\shownote{Publisher: Institute of Mathematical Statistics.}


\bibitem[\protect\citeauthoryear{Eger and Mathewson}{Eger and
  Mathewson}{2018}]%
        {eger_dairector_2018}
\bibfield{author}{\bibinfo{person}{Markus Eger} {and} \bibinfo{person}{Kory~W.
  Mathewson}.} \bibinfo{year}{2018}\natexlab{}.
\newblock \showarticletitle{{dAIrector}: {Automatic} {Story} {Beat}
  {Generation} through {Knowledge} {Synthesis}}.
\newblock \bibinfo{journal}{\emph{CoRR}}  \bibinfo{volume}{abs/1811.03423}
  (\bibinfo{year}{2018}).
\newblock
\urldef\tempurl%
\url{http://arxiv.org/abs/1811.03423}
\showURL{%
\tempurl}
\newblock
\shownote{arXiv: 1811.03423.}


\bibitem[\protect\citeauthoryear{Fairclough and Cunningham}{Fairclough and
  Cunningham}{2003}]%
        {fairclough_multiplayer_2003}
\bibfield{author}{\bibinfo{person}{C. Fairclough} {and} \bibinfo{person}{P.
  Cunningham}.} \bibinfo{year}{2003}\natexlab{}.
\newblock \showarticletitle{A {Multiplayer} {Case} {Based} {Story} {Engine}}.
\newblock
\urldef\tempurl%
\url{https://www.semanticscholar.org/paper/A-Multiplayer-Case-Based-Story-Engine-Fairclough-Cunningham/aea44aaafca25c3d4c1919a57258528d7dfbd798}
\showURL{%
\tempurl}


\bibitem[\protect\citeauthoryear{Fan, Lewis, and Dauphin}{Fan
  et~al\mbox{.}}{2018}]%
        {fan_hierarchical_2018}
\bibfield{author}{\bibinfo{person}{Angela Fan}, \bibinfo{person}{Mike Lewis},
  {and} \bibinfo{person}{Yann Dauphin}.} \bibinfo{year}{2018}\natexlab{}.
\newblock \bibinfo{title}{Hierarchical {Neural} {Story} {Generation}}.
\newblock
\newblock
\urldef\tempurl%
\url{https://doi.org/10.48550/arXiv.1805.04833}
\showDOI{\tempurl}
\newblock
\shownote{arXiv:1805.04833 [cs].}


\bibitem[\protect\citeauthoryear{García-Ortega, García-Sánchez, and
  Merelo-Guervós}{García-Ortega et~al\mbox{.}}{2020a}]%
        {garcia-ortega_startroper_2020}
\bibfield{author}{\bibinfo{person}{Rubén~Héctor García-Ortega},
  \bibinfo{person}{Pablo García-Sánchez}, {and} \bibinfo{person}{Juan~Julián
  Merelo-Guervós}.} \bibinfo{year}{2020}\natexlab{a}.
\newblock \showarticletitle{{StarTroper}, a film trope rating optimizer using
  machine learning and evolutionary algorithms}.
\newblock \bibinfo{journal}{\emph{Expert Systems}} \bibinfo{volume}{37},
  \bibinfo{number}{6} (\bibinfo{year}{2020}), \bibinfo{pages}{e12525}.
\newblock
\newblock
\shownote{Publisher: Wiley Online Library.}


\bibitem[\protect\citeauthoryear{García-Ortega, Merelo-Guervós, Sánchez, and
  Pitaru}{García-Ortega et~al\mbox{.}}{2018}]%
        {garcia-ortega_overview_2018}
\bibfield{author}{\bibinfo{person}{Rubén~H García-Ortega},
  \bibinfo{person}{Juan~J Merelo-Guervós}, \bibinfo{person}{Pablo~García
  Sánchez}, {and} \bibinfo{person}{Gad Pitaru}.}
  \bibinfo{year}{2018}\natexlab{}.
\newblock \showarticletitle{Overview of {PicTropes}, a film trope dataset}.
\newblock \bibinfo{journal}{\emph{arXiv preprint arXiv:1809.10959}}
  (\bibinfo{year}{2018}).
\newblock


\bibitem[\protect\citeauthoryear{García-Ortega, Sánchez, and
  Merelo-Guervós}{García-Ortega et~al\mbox{.}}{2020b}]%
        {garcia-ortega_tropes_2020}
\bibfield{author}{\bibinfo{person}{Rubén~Héctor García-Ortega},
  \bibinfo{person}{Pablo~García Sánchez}, {and} \bibinfo{person}{Juan~J
  Merelo-Guervós}.} \bibinfo{year}{2020}\natexlab{b}.
\newblock \showarticletitle{Tropes in films: an initial analysis}.
\newblock \bibinfo{journal}{\emph{arXiv preprint arXiv:2006.05380}}
  (\bibinfo{year}{2020}).
\newblock


\bibitem[\protect\citeauthoryear{García-Sánchez, Velez-Estevez, Merelo, and
  Cobo}{García-Sánchez et~al\mbox{.}}{2021}]%
        {garcia-sanchez_simpsons_2021}
\bibfield{author}{\bibinfo{person}{Pablo García-Sánchez},
  \bibinfo{person}{Antonio Velez-Estevez}, \bibinfo{person}{Juan~Julián
  Merelo}, {and} \bibinfo{person}{Manuel~Jesús Cobo}.}
  \bibinfo{year}{2021}\natexlab{}.
\newblock \showarticletitle{The {Simpsons} did it: {Exploring} the film trope
  space and its large scale structure}.
\newblock \bibinfo{journal}{\emph{Plos one}} \bibinfo{volume}{16},
  \bibinfo{number}{3} (\bibinfo{year}{2021}), \bibinfo{pages}{e0248881}.
\newblock
\newblock
\shownote{Publisher: Public Library of Science.}


\bibitem[\protect\citeauthoryear{Gervás, Díaz-Agudo, Peinado, and
  Hervás}{Gervás et~al\mbox{.}}{2005}]%
        {gervas_story_2005}
\bibfield{author}{\bibinfo{person}{Pablo Gervás}, \bibinfo{person}{Belén
  Díaz-Agudo}, \bibinfo{person}{Federico Peinado}, {and}
  \bibinfo{person}{Raquel Hervás}.} \bibinfo{year}{2005}\natexlab{}.
\newblock \showarticletitle{Story plot generation based on {CBR}}.
\newblock \bibinfo{journal}{\emph{Knowledge-Based Systems}}
  \bibinfo{volume}{18}, \bibinfo{number}{4} (\bibinfo{date}{Aug.}
  \bibinfo{year}{2005}), \bibinfo{pages}{235--242}.
\newblock
\showISSN{0950-7051}
\urldef\tempurl%
\url{https://doi.org/10.1016/j.knosys.2004.10.011}
\showDOI{\tempurl}


\bibitem[\protect\citeauthoryear{Gonsalves}{Gonsalves}{2023}]%
        {gonsalves_using_2023}
\bibfield{author}{\bibinfo{person}{Robert~A. Gonsalves}.}
  \bibinfo{year}{2023}\natexlab{}.
\newblock \bibinfo{title}{Using {ChatGPT} as a {Creative} {Writing} {Partner}
  — {Part} 1: {Prose}}.
\newblock
\newblock
\urldef\tempurl%
\url{https://towardsdatascience.com/using-chatgpt-as-a-creative-writing-partner-part-1-prose-dc9a9994d41f}
\showURL{%
\tempurl}


\bibitem[\protect\citeauthoryear{Grasbon and Braun}{Grasbon and Braun}{2001}]%
        {grasbon_morphological_2001}
\bibfield{author}{\bibinfo{person}{D. Grasbon} {and} \bibinfo{person}{N.
  Braun}.} \bibinfo{year}{2001}\natexlab{}.
\newblock \showarticletitle{A {Morphological} {Approach} to {Interactive}
  {Storytelling}}.
\newblock  (\bibinfo{year}{2001}).
\newblock
\urldef\tempurl%
\url{https://publica.fraunhofer.de/handle/publica/338099}
\showURL{%
\tempurl}


\bibitem[\protect\citeauthoryear{Guarneri, Ripamonti, Tissoni, Trubian,
  Maggiorini, and Gadia}{Guarneri et~al\mbox{.}}{2017}]%
        {guarneri_ghost_2017}
\bibfield{author}{\bibinfo{person}{Andrea Guarneri}, \bibinfo{person}{Laura
  Ripamonti}, \bibinfo{person}{Francesco Tissoni}, \bibinfo{person}{Marco
  Trubian}, \bibinfo{person}{Dario Maggiorini}, {and} \bibinfo{person}{Davide
  Gadia}.} \bibinfo{year}{2017}\natexlab{}.
\newblock \showarticletitle{{GHOST}: a {GHOst} {STory}-writer}.
  \bibinfo{pages}{1--9}.
\newblock
\urldef\tempurl%
\url{https://doi.org/10.1145/3125571.3125580}
\showDOI{\tempurl}


\bibitem[\protect\citeauthoryear{Harper and Konstan}{Harper and
  Konstan}{2015}]%
        {harper_movielens_2015}
\bibfield{author}{\bibinfo{person}{F.~Maxwell Harper} {and}
  \bibinfo{person}{Joseph~A. Konstan}.} \bibinfo{year}{2015}\natexlab{}.
\newblock \showarticletitle{The {MovieLens} {Datasets}: {History} and
  {Context}}.
\newblock \bibinfo{journal}{\emph{ACM Transactions on Interactive Intelligent
  Systems}} \bibinfo{volume}{5}, \bibinfo{number}{4} (\bibinfo{date}{Dec.}
  \bibinfo{year}{2015}), \bibinfo{pages}{19:1--19:19}.
\newblock
\showISSN{2160-6455}
\urldef\tempurl%
\url{https://doi.org/10.1145/2827872}
\showDOI{\tempurl}


\bibitem[\protect\citeauthoryear{Harris}{Harris}{2017}]%
        {harris_periodic_2017}
\bibfield{author}{\bibinfo{person}{James Harris}.}
  \bibinfo{year}{2017}\natexlab{}.
\newblock \bibinfo{title}{The {Periodic} {Table} of {Storytelling}}.
\newblock
\newblock
\urldef\tempurl%
\url{https://jamesharris.design/periodic/}
\showURL{%
\tempurl}


\bibitem[\protect\citeauthoryear{Horswill}{Horswill}{2021}]%
        {horswill_dear_2021}
\bibfield{author}{\bibinfo{person}{Ian Horswill}.}
  \bibinfo{year}{2021}\natexlab{}.
\newblock \showarticletitle{Dear {Leader}’s {Happy} {Story} {Time}: {A}
  {Party} {Game} {Based} on {Automated} {Story} {Generation}}.
\newblock \bibinfo{journal}{\emph{Proceedings of the AAAI Conference on
  Artificial Intelligence and Interactive Digital Entertainment}}
  \bibinfo{volume}{12}, \bibinfo{number}{2} (\bibinfo{date}{June}
  \bibinfo{year}{2021}), \bibinfo{pages}{39--45}.
\newblock
\urldef\tempurl%
\url{https://doi.org/10.1609/aiide.v12i2.12902}
\showDOI{\tempurl}


\bibitem[\protect\citeauthoryear{Huang, Huang, and Huang}{Huang
  et~al\mbox{.}}{2020a}]%
        {huang_heteroglossia_2020}
\bibfield{author}{\bibinfo{person}{Chieh-Yang Huang},
  \bibinfo{person}{Shih-Hong Huang}, {and} \bibinfo{person}{Ting-Hao~Kenneth
  Huang}.} \bibinfo{year}{2020}\natexlab{a}.
\newblock \showarticletitle{Heteroglossia: {In}-{Situ} {Story} {Ideation} with
  the {Crowd}}. In \bibinfo{booktitle}{\emph{Proceedings of the 2020 {CHI}
  {Conference} on {Human} {Factors} in {Computing} {Systems}}}
  \emph{(\bibinfo{series}{{CHI} '20})}. \bibinfo{publisher}{Association for
  Computing Machinery}, \bibinfo{address}{New York, NY, USA},
  \bibinfo{pages}{1--12}.
\newblock
\showISBNx{978-1-4503-6708-0}
\urldef\tempurl%
\url{https://doi.org/10.1145/3313831.3376715}
\showDOI{\tempurl}


\bibitem[\protect\citeauthoryear{Huang, Zhang, Elachqar, and Cheng}{Huang
  et~al\mbox{.}}{2020b}]%
        {huang_inset_2020}
\bibfield{author}{\bibinfo{person}{Yichen Huang}, \bibinfo{person}{Yizhe
  Zhang}, \bibinfo{person}{Oussama Elachqar}, {and} \bibinfo{person}{Yu
  Cheng}.} \bibinfo{year}{2020}\natexlab{b}.
\newblock \showarticletitle{{INSET}: {Sentence} {Infilling} with
  {INter}-{SEntential} {Transformer}}. In \bibinfo{booktitle}{\emph{Proceedings
  of the 58th {Annual} {Meeting} of the {Association} for {Computational}
  {Linguistics}}}. \bibinfo{publisher}{Association for Computational
  Linguistics}, \bibinfo{address}{Online}, \bibinfo{pages}{2502--2515}.
\newblock
\urldef\tempurl%
\url{https://doi.org/10.18653/v1/2020.acl-main.226}
\showDOI{\tempurl}


\bibitem[\protect\citeauthoryear{Ippolito, Grangier, Callison-Burch, and
  Eck}{Ippolito et~al\mbox{.}}{2019}]%
        {ippolito_unsupervised_2019}
\bibfield{author}{\bibinfo{person}{Daphne Ippolito}, \bibinfo{person}{David
  Grangier}, \bibinfo{person}{Chris Callison-Burch}, {and}
  \bibinfo{person}{Douglas Eck}.} \bibinfo{year}{2019}\natexlab{}.
\newblock \showarticletitle{Unsupervised {Hierarchical} {Story} {Infilling}}.
  In \bibinfo{booktitle}{\emph{Proceedings of the {First} {Workshop} on
  {Narrative} {Understanding}}}. \bibinfo{publisher}{Association for
  Computational Linguistics}, \bibinfo{address}{Minneapolis, Minnesota},
  \bibinfo{pages}{37--43}.
\newblock
\urldef\tempurl%
\url{https://doi.org/10.18653/v1/W19-2405}
\showDOI{\tempurl}


\bibitem[\protect\citeauthoryear{Johnson}{Johnson}{2018}]%
        {johnson_scaling_2018}
\bibfield{author}{\bibinfo{person}{Alister Johnson}.}
  \bibinfo{year}{2018}\natexlab{}.
\newblock \showarticletitle{Scaling {Collaborative} {Filtering} with {PETSc}}.
  In \bibinfo{booktitle}{\emph{2018 {IEEE} {International} {Conference} on
  {Big} {Data} ({Big} {Data})}}. \bibinfo{pages}{4237--4244}.
\newblock
\urldef\tempurl%
\url{https://doi.org/10.1109/BigData.2018.8622202}
\showDOI{\tempurl}


\bibitem[\protect\citeauthoryear{Kiesel and Grimnes}{Kiesel and
  Grimnes}{2010}]%
        {kiesel_dbtropes-linked_2010}
\bibfield{author}{\bibinfo{person}{Malte Kiesel} {and}
  \bibinfo{person}{Gunnar~Aastrand Grimnes}.} \bibinfo{year}{2010}\natexlab{}.
\newblock \showarticletitle{{DBTropes}-a {Linked} {Data} {Wrapper} {Approach}
  {Incorporating} {Community} {Feedback}.}. In \bibinfo{booktitle}{\emph{{EKAW}
  ({Posters} and {Demos})}}. \bibinfo{publisher}{Citeseer}.
\newblock


\bibitem[\protect\citeauthoryear{Kim, Sterman, Cohen, and Bernstein}{Kim
  et~al\mbox{.}}{2017}]%
        {kim_mechanical_2017}
\bibfield{author}{\bibinfo{person}{Joy Kim}, \bibinfo{person}{Sarah Sterman},
  \bibinfo{person}{Allegra Argent~Beal Cohen}, {and}
  \bibinfo{person}{Michael~S. Bernstein}.} \bibinfo{year}{2017}\natexlab{}.
\newblock \showarticletitle{Mechanical {Novel}: {Crowdsourcing} {Complex}
  {Work} through {Reflection} and {Revision}}. In
  \bibinfo{booktitle}{\emph{Proceedings of the 2017 {ACM} {Conference} on
  {Computer} {Supported} {Cooperative} {Work} and {Social} {Computing}}}
  \emph{(\bibinfo{series}{{CSCW} '17})}. \bibinfo{publisher}{Association for
  Computing Machinery}, \bibinfo{address}{New York, NY, USA},
  \bibinfo{pages}{233--245}.
\newblock
\showISBNx{978-1-4503-4335-0}
\urldef\tempurl%
\url{https://doi.org/10.1145/2998181.2998196}
\showDOI{\tempurl}


\bibitem[\protect\citeauthoryear{Koch, Lucero, Hegemann, and Oulasvirta}{Koch
  et~al\mbox{.}}{2019}]%
        {koch_may_2019}
\bibfield{author}{\bibinfo{person}{Janin Koch}, \bibinfo{person}{Andr\'{e}s
  Lucero}, \bibinfo{person}{Lena Hegemann}, {and} \bibinfo{person}{Antti
  Oulasvirta}.} \bibinfo{year}{2019}\natexlab{}.
\newblock \showarticletitle{May AI? Design Ideation with Cooperative Contextual
  Bandits}. In \bibinfo{booktitle}{\emph{Proceedings of the 2019 CHI Conference
  on Human Factors in Computing Systems}} (Glasgow, Scotland Uk)
  \emph{(\bibinfo{series}{CHI '19})}. \bibinfo{publisher}{Association for
  Computing Machinery}, \bibinfo{address}{New York, NY, USA},
  \bibinfo{pages}{1–12}.
\newblock
\showISBNx{9781450359702}
\urldef\tempurl%
\url{https://doi.org/10.1145/3290605.3300863}
\showDOI{\tempurl}


\bibitem[\protect\citeauthoryear{Koch, Taffin, Beaudouin-Lafon, Laine, Lucero,
  and Mackay}{Koch et~al\mbox{.}}{2020a}]%
        {koch_imagesense_2020}
\bibfield{author}{\bibinfo{person}{Janin Koch}, \bibinfo{person}{Nicolas
  Taffin}, \bibinfo{person}{Michel Beaudouin-Lafon}, \bibinfo{person}{Markku
  Laine}, \bibinfo{person}{Andr\'{e}s Lucero}, {and} \bibinfo{person}{Wendy~E.
  Mackay}.} \bibinfo{year}{2020}\natexlab{a}.
\newblock \showarticletitle{ImageSense: An Intelligent Collaborative Ideation
  Tool to Support Diverse Human-Computer Partnerships}.
\newblock \bibinfo{journal}{\emph{Proc. ACM Hum.-Comput. Interact.}}
  \bibinfo{volume}{4}, \bibinfo{number}{CSCW1}, Article \bibinfo{articleno}{45}
  (\bibinfo{date}{may} \bibinfo{year}{2020}), \bibinfo{numpages}{27}~pages.
\newblock
\urldef\tempurl%
\url{https://doi.org/10.1145/3392850}
\showDOI{\tempurl}


\bibitem[\protect\citeauthoryear{Koch, Taffin, Lucero, and Mackay}{Koch
  et~al\mbox{.}}{2020b}]%
        {koch_semanticcollage_2020}
\bibfield{author}{\bibinfo{person}{Janin Koch}, \bibinfo{person}{Nicolas
  Taffin}, \bibinfo{person}{Andr\'{e}s Lucero}, {and} \bibinfo{person}{Wendy~E.
  Mackay}.} \bibinfo{year}{2020}\natexlab{b}.
\newblock \showarticletitle{SemanticCollage: Enriching Digital Mood Board
  Design with Semantic Labels}. In \bibinfo{booktitle}{\emph{Proceedings of the
  2020 ACM Designing Interactive Systems Conference}} (Eindhoven, Netherlands)
  \emph{(\bibinfo{series}{DIS '20})}. \bibinfo{publisher}{Association for
  Computing Machinery}, \bibinfo{address}{New York, NY, USA},
  \bibinfo{pages}{407–418}.
\newblock
\showISBNx{9781450369749}
\urldef\tempurl%
\url{https://doi.org/10.1145/3357236.3395494}
\showDOI{\tempurl}


\bibitem[\protect\citeauthoryear{Lakoff}{Lakoff}{1972}]%
        {lakoff_structural_1972}
\bibfield{author}{\bibinfo{person}{George Lakoff}.}
  \bibinfo{year}{1972}\natexlab{}.
\newblock \showarticletitle{Structural {Complexity} in {Fairy} {Tales}}.
\newblock  (\bibinfo{year}{1972}).
\newblock
\urldef\tempurl%
\url{https://escholarship.org/uc/item/6h38w8jc}
\showURL{%
\tempurl}


\bibitem[\protect\citeauthoryear{Lebowitz}{Lebowitz}{1984}]%
        {lebowitz_creating_1984}
\bibfield{author}{\bibinfo{person}{Michael Lebowitz}.}
  \bibinfo{year}{1984}\natexlab{}.
\newblock \showarticletitle{Creating characters in a story-telling universe}.
\newblock \bibinfo{journal}{\emph{Poetics}} \bibinfo{volume}{13},
  \bibinfo{number}{3} (\bibinfo{date}{June} \bibinfo{year}{1984}),
  \bibinfo{pages}{171--194}.
\newblock
\showISSN{0304-422X}
\urldef\tempurl%
\url{https://doi.org/10.1016/0304-422X(84)90001-9}
\showDOI{\tempurl}


\bibitem[\protect\citeauthoryear{Meehan}{Meehan}{1977}]%
        {meehan_tale-spin_1977}
\bibfield{author}{\bibinfo{person}{James~R. Meehan}.}
  \bibinfo{year}{1977}\natexlab{}.
\newblock \showarticletitle{{TALE}-{SPIN}, an interactive program that writes
  stories}. In \bibinfo{booktitle}{\emph{Proceedings of the 5th international
  joint conference on {Artificial} intelligence - {Volume} 1}}
  \emph{(\bibinfo{series}{{IJCAI}'77})}. \bibinfo{publisher}{Morgan Kaufmann
  Publishers Inc.}, \bibinfo{address}{San Francisco, CA, USA},
  \bibinfo{pages}{91--98}.
\newblock


\bibitem[\protect\citeauthoryear{Mellina and Svetlichnaya}{Mellina and
  Svetlichnaya}{2011}]%
        {mellina_trope_2011}
\bibfield{author}{\bibinfo{person}{Clayton Mellina} {and}
  \bibinfo{person}{Stacey Svetlichnaya}.} \bibinfo{year}{2011}\natexlab{}.
\newblock \bibinfo{title}{Trope propagation in the cultural space}.
\newblock
\newblock


\bibitem[\protect\citeauthoryear{Mirowski, Mathewson, Pittman, and
  Evans}{Mirowski et~al\mbox{.}}{2023}]%
        {mirowski_co_2023}
\bibfield{author}{\bibinfo{person}{Piotr Mirowski}, \bibinfo{person}{Kory~W.
  Mathewson}, \bibinfo{person}{Jaylen Pittman}, {and} \bibinfo{person}{Richard
  Evans}.} \bibinfo{year}{2023}\natexlab{}.
\newblock \showarticletitle{Co-Writing Screenplays and Theatre Scripts with
  Language Models: Evaluation by Industry Professionals}. In
  \bibinfo{booktitle}{\emph{Proceedings of the 2023 CHI Conference on Human
  Factors in Computing Systems}} (Hamburg, Germany) \emph{(\bibinfo{series}{CHI
  '23})}. \bibinfo{publisher}{Association for Computing Machinery},
  \bibinfo{address}{New York, NY, USA}, Article \bibinfo{articleno}{355},
  \bibinfo{numpages}{34}~pages.
\newblock
\showISBNx{9781450394215}
\urldef\tempurl%
\url{https://doi.org/10.1145/3544548.3581225}
\showDOI{\tempurl}


\bibitem[\protect\citeauthoryear{Mitchell, Wu, Zaldivar, Barnes, Vasserman,
  Hutchinson, Spitzer, Raji, and Gebru}{Mitchell et~al\mbox{.}}{2019}]%
        {mitchell_model_2019}
\bibfield{author}{\bibinfo{person}{Margaret Mitchell}, \bibinfo{person}{Simone
  Wu}, \bibinfo{person}{Andrew Zaldivar}, \bibinfo{person}{Parker Barnes},
  \bibinfo{person}{Lucy Vasserman}, \bibinfo{person}{Ben Hutchinson},
  \bibinfo{person}{Elena Spitzer}, \bibinfo{person}{Inioluwa~Deborah Raji},
  {and} \bibinfo{person}{Timnit Gebru}.} \bibinfo{year}{2019}\natexlab{}.
\newblock \showarticletitle{Model Cards for Model Reporting}. In
  \bibinfo{booktitle}{\emph{Proceedings of the Conference on Fairness,
  Accountability, and Transparency}} (Atlanta, GA, USA)
  \emph{(\bibinfo{series}{FAT* '19})}. \bibinfo{publisher}{Association for
  Computing Machinery}, \bibinfo{address}{New York, NY, USA},
  \bibinfo{pages}{220–229}.
\newblock
\showISBNx{9781450361255}
\urldef\tempurl%
\url{https://doi.org/10.1145/3287560.3287596}
\showDOI{\tempurl}


\bibitem[\protect\citeauthoryear{Pedregosa, Varoquaux, Gramfort, Michel,
  Thirion, Grisel, Blondel, Prettenhofer, Weiss, Dubourg, Vanderplas, Passos,
  Cournapeau, Brucher, Perrot, and Duchesnay}{Pedregosa et~al\mbox{.}}{2011}]%
        {scikit-learn}
\bibfield{author}{\bibinfo{person}{F. Pedregosa}, \bibinfo{person}{G.
  Varoquaux}, \bibinfo{person}{A. Gramfort}, \bibinfo{person}{V. Michel},
  \bibinfo{person}{B. Thirion}, \bibinfo{person}{O. Grisel},
  \bibinfo{person}{M. Blondel}, \bibinfo{person}{P. Prettenhofer},
  \bibinfo{person}{R. Weiss}, \bibinfo{person}{V. Dubourg}, \bibinfo{person}{J.
  Vanderplas}, \bibinfo{person}{A. Passos}, \bibinfo{person}{D. Cournapeau},
  \bibinfo{person}{M. Brucher}, \bibinfo{person}{M. Perrot}, {and}
  \bibinfo{person}{E. Duchesnay}.} \bibinfo{year}{2011}\natexlab{}.
\newblock \showarticletitle{Scikit-learn: Machine Learning in {P}ython}.
\newblock \bibinfo{journal}{\emph{Journal of Machine Learning Research}}
  \bibinfo{volume}{12} (\bibinfo{year}{2011}), \bibinfo{pages}{2825--2830}.
\newblock


\bibitem[\protect\citeauthoryear{Propp}{Propp}{1928}]%
        {propp_morphology_1928}
\bibfield{author}{\bibinfo{person}{Vladimir Propp}.}
  \bibinfo{year}{1928}\natexlab{}.
\newblock \bibinfo{booktitle}{\emph{Morphology of the {Folktale}}}.
  Vol.~\bibinfo{volume}{9}.
\newblock \bibinfo{publisher}{University of Texas Press}.
\newblock


\bibitem[\protect\citeauthoryear{Purdy, Wang, He, and Riedl}{Purdy
  et~al\mbox{.}}{2018}]%
        {purdy_predicting_2018}
\bibfield{author}{\bibinfo{person}{Christopher Purdy}, \bibinfo{person}{Xinyu
  Wang}, \bibinfo{person}{Larry He}, {and} \bibinfo{person}{Mark Riedl}.}
  \bibinfo{year}{2018}\natexlab{}.
\newblock \showarticletitle{Predicting generated story quality with
  quantitative measures}. In \bibinfo{booktitle}{\emph{Proceedings of the
  {Fourteenth} {AAAI} {Conference} on {Artificial} {Intelligence} and
  {Interactive} {Digital} {Entertainment}}}
  \emph{(\bibinfo{series}{{AIIDE}'18})}. \bibinfo{publisher}{AAAI Press},
  \bibinfo{address}{Edmonton, Alberta, Canada}, \bibinfo{pages}{95--101}.
\newblock
\showISBNx{978-1-57735-804-6}


\bibitem[\protect\citeauthoryear{PÉrez and Sharples}{PÉrez and
  Sharples}{2001}]%
        {perez_mexica_2001}
\bibfield{author}{\bibinfo{person}{Rafael PÉrez~Ý PÉrez} {and}
  \bibinfo{person}{Mike Sharples}.} \bibinfo{year}{2001}\natexlab{}.
\newblock \showarticletitle{{MEXICA}: {A} computer model of a cognitive account
  of creative writing}.
\newblock \bibinfo{journal}{\emph{Journal of Experimental \& Theoretical
  Artificial Intelligence}} \bibinfo{volume}{13}, \bibinfo{number}{2}
  (\bibinfo{date}{April} \bibinfo{year}{2001}), \bibinfo{pages}{119--139}.
\newblock
\showISSN{0952-813X}
\urldef\tempurl%
\url{https://doi.org/10.1080/09528130010029820}
\showDOI{\tempurl}
\newblock
\shownote{Publisher: Taylor \& Francis \_eprint:
  https://doi.org/10.1080/09528130010029820.}


\bibitem[\protect\citeauthoryear{Riedl}{Riedl}{2009}]%
        {riedl_vignette-based_2009}
\bibfield{author}{\bibinfo{person}{Mark Riedl}.}
  \bibinfo{year}{2009}\natexlab{}.
\newblock \showarticletitle{Vignette-{Based} {Story} {Planning}: {Creativity}
  {Through} {Exploration} and {Retrieval}}.
\newblock \bibinfo{journal}{\emph{Proceedings of the International Joint
  Workshop on Computational Creativity 2008}} (\bibinfo{date}{Jan.}
  \bibinfo{year}{2009}).
\newblock


\bibitem[\protect\citeauthoryear{Riedl and Young}{Riedl and Young}{2010}]%
        {riedl_narrative_2010}
\bibfield{author}{\bibinfo{person}{Mark~O. Riedl} {and}
  \bibinfo{person}{R.~Michael Young}.} \bibinfo{year}{2010}\natexlab{}.
\newblock \showarticletitle{Narrative planning: balancing plot and character}.
\newblock \bibinfo{journal}{\emph{Journal of Artificial Intelligence Research}}
  \bibinfo{volume}{39}, \bibinfo{number}{1} (\bibinfo{date}{Sept.}
  \bibinfo{year}{2010}), \bibinfo{pages}{217--268}.
\newblock
\showISSN{1076-9757}


\bibitem[\protect\citeauthoryear{Roemmele, Gordon, and Swanson}{Roemmele
  et~al\mbox{.}}{2017}]%
        {roemmele_evaluating_2017}
\bibfield{author}{\bibinfo{person}{Melissa Roemmele}, \bibinfo{person}{A.
  Gordon}, {and} \bibinfo{person}{R. Swanson}.}
  \bibinfo{year}{2017}\natexlab{}.
\newblock \showarticletitle{Evaluating {Story} {Generation} {Systems} {Using}
  {Automated} {Linguistic} {Analyses}}.
\newblock
\urldef\tempurl%
\url{https://www.semanticscholar.org/paper/Evaluating-Story-Generation-Systems-Using-Automated-Roemmele-Gordon/cf222683dd06990d76da48612c4dbe72d62f968d}
\showURL{%
\tempurl}


\bibitem[\protect\citeauthoryear{Roemmele and Gordon}{Roemmele and
  Gordon}{2015}]%
        {roemmele_creative_2015}
\bibfield{author}{\bibinfo{person}{Melissa Roemmele} {and}
  \bibinfo{person}{Andrew~S. Gordon}.} \bibinfo{year}{2015}\natexlab{}.
\newblock \showarticletitle{Creative {Help}: {A} {Story} {Writing}
  {Assistant}}. In \bibinfo{booktitle}{\emph{Interactive {Storytelling}}}
  \emph{(\bibinfo{series}{Lecture {Notes} in {Computer} {Science}})},
  \bibfield{editor}{\bibinfo{person}{Henrik Schoenau-Fog},
  \bibinfo{person}{Luis~Emilio Bruni}, \bibinfo{person}{Sandy Louchart}, {and}
  \bibinfo{person}{Sarune Baceviciute}} (Eds.). \bibinfo{publisher}{Springer
  International Publishing}, \bibinfo{address}{Cham}, \bibinfo{pages}{81--92}.
\newblock
\showISBNx{978-3-319-27036-4}
\urldef\tempurl%
\url{https://doi.org/10.1007/978-3-319-27036-4_8}
\showDOI{\tempurl}


\bibitem[\protect\citeauthoryear{Rumelhart}{Rumelhart}{1975}]%
        {rumelhart_notes_1975}
\bibfield{author}{\bibinfo{person}{David~E. Rumelhart}.}
  \bibinfo{year}{1975}\natexlab{}.
\newblock \showarticletitle{{NOTES} {ON} {A} {SCHEMA} {FOR} {STORIES}}.
\newblock In \bibinfo{booktitle}{\emph{Representation and {Understanding}}},
  \bibfield{editor}{\bibinfo{person}{DANIEL~G. Bobrow} {and}
  \bibinfo{person}{ALLAN Collins}} (Eds.). \bibinfo{publisher}{Morgan
  Kaufmann}, \bibinfo{address}{San Diego}, \bibinfo{pages}{211--236}.
\newblock
\showISBNx{978-0-12-108550-6}
\urldef\tempurl%
\url{https://doi.org/10.1016/B978-0-12-108550-6.50013-6}
\showDOI{\tempurl}


\bibitem[\protect\citeauthoryear{Sagarkar, Wieting, Tu, and Gimpel}{Sagarkar
  et~al\mbox{.}}{2018}]%
        {sagarkar_quality_2018}
\bibfield{author}{\bibinfo{person}{Manasvi Sagarkar}, \bibinfo{person}{John
  Wieting}, \bibinfo{person}{Lifu Tu}, {and} \bibinfo{person}{Kevin Gimpel}.}
  \bibinfo{year}{2018}\natexlab{}.
\newblock \showarticletitle{Quality {Signals} in {Generated} {Stories}}. In
  \bibinfo{booktitle}{\emph{Proceedings of the {Seventh} {Joint} {Conference}
  on {Lexical} and {Computational} {Semantics}}}.
  \bibinfo{publisher}{Association for Computational Linguistics},
  \bibinfo{address}{New Orleans, Louisiana}, \bibinfo{pages}{192--202}.
\newblock
\urldef\tempurl%
\url{https://doi.org/10.18653/v1/S18-2024}
\showDOI{\tempurl}


\bibitem[\protect\citeauthoryear{Smith, Joshi, Huet, Hsu, and Cota}{Smith
  et~al\mbox{.}}{2017}]%
        {smith_harnessing_2017}
\bibfield{author}{\bibinfo{person}{John~R Smith}, \bibinfo{person}{Dhiraj
  Joshi}, \bibinfo{person}{Benoit Huet}, \bibinfo{person}{Winston Hsu}, {and}
  \bibinfo{person}{Jozef Cota}.} \bibinfo{year}{2017}\natexlab{}.
\newblock \showarticletitle{Harnessing ai for augmenting creativity:
  {Application} to movie trailer creation}. In
  \bibinfo{booktitle}{\emph{Proceedings of the 25th {ACM} international
  conference on {Multimedia}}}. \bibinfo{pages}{1799--1808}.
\newblock


\bibitem[\protect\citeauthoryear{Su, Shen, Tsai, Cheng, Wang, and Hsu}{Su
  et~al\mbox{.}}{2021}]%
        {su_truman_2021}
\bibfield{author}{\bibinfo{person}{Hung-Ting Su}, \bibinfo{person}{Po-Wei
  Shen}, \bibinfo{person}{Bing-Chen Tsai}, \bibinfo{person}{Wen-Feng Cheng},
  \bibinfo{person}{Ke-Jyun Wang}, {and} \bibinfo{person}{Winston~H. Hsu}.}
  \bibinfo{year}{2021}\natexlab{}.
\newblock \showarticletitle{{TrUMAn}: {Trope} {Understanding} in {Movies} and
  {Animations}}. In \bibinfo{booktitle}{\emph{Proceedings of the 30th {ACM}
  {International} {Conference} on {Information} \& {Knowledge} {Management}}}
  \emph{(\bibinfo{series}{{CIKM} '21})}. \bibinfo{publisher}{Association for
  Computing Machinery}, \bibinfo{address}{New York, NY, USA},
  \bibinfo{pages}{4594--4603}.
\newblock
\showISBNx{978-1-4503-8446-9}
\urldef\tempurl%
\url{https://doi.org/10.1145/3459637.3482018}
\showDOI{\tempurl}


\bibitem[\protect\citeauthoryear{Sun, Zhao, Manjunatha, Jain, Morariu,
  Dernoncourt, Srinivasan, and Iyyer}{Sun et~al\mbox{.}}{2021}]%
        {sun_iga_2021}
\bibfield{author}{\bibinfo{person}{Simeng Sun}, \bibinfo{person}{Wenlong Zhao},
  \bibinfo{person}{Varun Manjunatha}, \bibinfo{person}{Rajiv Jain},
  \bibinfo{person}{Vlad Morariu}, \bibinfo{person}{Franck Dernoncourt},
  \bibinfo{person}{Balaji~Vasan Srinivasan}, {and} \bibinfo{person}{Mohit
  Iyyer}.} \bibinfo{year}{2021}\natexlab{}.
\newblock \showarticletitle{{IGA}: {An} {Intent}-{Guided} {Authoring}
  {Assistant}}. In \bibinfo{booktitle}{\emph{Proceedings of the 2021
  {Conference} on {Empirical} {Methods} in {Natural} {Language} {Processing}}}.
  \bibinfo{publisher}{Association for Computational Linguistics},
  \bibinfo{address}{Online and Punta Cana, Dominican Republic},
  \bibinfo{pages}{5972--5985}.
\newblock
\urldef\tempurl%
\url{https://doi.org/10.18653/v1/2021.emnlp-main.483}
\showDOI{\tempurl}


\bibitem[\protect\citeauthoryear{Swanson and Gordon}{Swanson and
  Gordon}{2012}]%
        {swanson_say_2012}
\bibfield{author}{\bibinfo{person}{Reid Swanson} {and}
  \bibinfo{person}{Andrew~S. Gordon}.} \bibinfo{year}{2012}\natexlab{}.
\newblock \showarticletitle{Say {Anything}: {Using} {Textual} {Case}-{Based}
  {Reasoning} to {Enable} {Open}-{Domain} {Interactive} {Storytelling}}.
\newblock \bibinfo{journal}{\emph{ACM Transactions on Interactive Intelligent
  Systems}} \bibinfo{volume}{2}, \bibinfo{number}{3} (\bibinfo{date}{Sept.}
  \bibinfo{year}{2012}), \bibinfo{pages}{16:1--16:35}.
\newblock
\showISSN{2160-6455}
\urldef\tempurl%
\url{https://doi.org/10.1145/2362394.2362398}
\showDOI{\tempurl}


\bibitem[\protect\citeauthoryear{Turner}{Turner}{[n.d.]}]%
        {turner_minstrel_nodate}
\bibfield{author}{\bibinfo{person}{Scott~R. Turner}.}
  \bibinfo{year}{[n.d.]}\natexlab{}.
\newblock \emph{\bibinfo{title}{{MINSTREL}: {A} computer model of creativity
  and storytelling}}.
\newblock Ph.{D}. \bibinfo{school}{University of California, Los Angeles},
  \bibinfo{address}{United States -- California}.
\newblock
\urldef\tempurl%
\url{https://www.proquest.com/docview/304049508/abstract/7A5295B0C69E46D3PQ/1}
\showURL{%
\tempurl}
\newblock
\shownote{ISBN: 9798209134299.}


\bibitem[\protect\citeauthoryear{Wang, Durrett, and Erk}{Wang
  et~al\mbox{.}}{2020}]%
        {wang_narrative_2020}
\bibfield{author}{\bibinfo{person}{Su Wang}, \bibinfo{person}{Greg Durrett},
  {and} \bibinfo{person}{Katrin Erk}.} \bibinfo{year}{2020}\natexlab{}.
\newblock \bibinfo{title}{Narrative {Interpolation} for {Generating} and
  {Understanding} {Stories}}.
\newblock
\newblock
\urldef\tempurl%
\url{https://doi.org/10.48550/arXiv.2008.07466}
\showDOI{\tempurl}
\newblock
\shownote{arXiv:2008.07466 [cs].}


\bibitem[\protect\citeauthoryear{Xu, Patwary, Shoeybi, Puri, Fung, Anandkumar,
  and Catanzaro}{Xu et~al\mbox{.}}{2020}]%
        {xu_megatron-cntrl_2020}
\bibfield{author}{\bibinfo{person}{Peng Xu}, \bibinfo{person}{Mostofa Patwary},
  \bibinfo{person}{Mohammad Shoeybi}, \bibinfo{person}{Raul Puri},
  \bibinfo{person}{Pascale Fung}, \bibinfo{person}{Anima Anandkumar}, {and}
  \bibinfo{person}{Bryan Catanzaro}.} \bibinfo{year}{2020}\natexlab{}.
\newblock \showarticletitle{{MEGATRON}-{CNTRL}: {Controllable} {Story}
  {Generation} with {External} {Knowledge} {Using} {Large}-{Scale} {Language}
  {Models}}. In \bibinfo{booktitle}{\emph{Proceedings of the 2020 {Conference}
  on {Empirical} {Methods} in {Natural} {Language} {Processing} ({EMNLP})}}.
  \bibinfo{publisher}{Association for Computational Linguistics},
  \bibinfo{address}{Online}, \bibinfo{pages}{2831--2845}.
\newblock
\urldef\tempurl%
\url{https://doi.org/10.18653/v1/2020.emnlp-main.226}
\showDOI{\tempurl}


\bibitem[\protect\citeauthoryear{Yuan, Coenen, Reif, and Ippolito}{Yuan
  et~al\mbox{.}}{2022}]%
        {yuan_wordcraft_2022}
\bibfield{author}{\bibinfo{person}{Ann Yuan}, \bibinfo{person}{Andy Coenen},
  \bibinfo{person}{Emily Reif}, {and} \bibinfo{person}{Daphne Ippolito}.}
  \bibinfo{year}{2022}\natexlab{}.
\newblock \showarticletitle{Wordcraft: {Story} {Writing} {With} {Large}
  {Language} {Models}}. In \bibinfo{booktitle}{\emph{27th {International}
  {Conference} on {Intelligent} {User} {Interfaces}}}
  \emph{(\bibinfo{series}{{IUI} '22})}. \bibinfo{publisher}{Association for
  Computing Machinery}, \bibinfo{address}{New York, NY, USA},
  \bibinfo{pages}{841--852}.
\newblock
\showISBNx{978-1-4503-9144-3}
\urldef\tempurl%
\url{https://doi.org/10.1145/3490099.3511105}
\showDOI{\tempurl}


\bibitem[\protect\citeauthoryear{Äijälä and {others}}{Äijälä and
  {others}}{2020}]%
        {aijala_using_2020}
\bibfield{author}{\bibinfo{person}{Cecilia Äijälä} {and}
  \bibinfo{person}{{others}}.} \bibinfo{year}{2020}\natexlab{}.
\newblock \showarticletitle{Using {Film}-{Trope} {Connections} for {Clustering}
  {Similar} {Movies}}.
\newblock  (\bibinfo{year}{2020}).
\newblock
\newblock
\shownote{Publisher: Helsingin yliopisto.}


\end{thebibliography}

%\end{sloppypar}
\end{document}